\documentclass[pdflatex,sn-mathphys-num]{sn-jnl}

\usepackage{graphicx}%
\usepackage{multirow}%
\usepackage{amsmath,amssymb,amsfonts}%
\usepackage{amsthm}%
\usepackage{mathrsfs}%
\usepackage[title]{appendix}%
\usepackage{xcolor}%
\usepackage{textcomp}%
\usepackage{manyfoot}%
\usepackage{booktabs}%
\usepackage{algorithm}%
\usepackage{algorithmicx}%
\usepackage{algpseudocode}%
\usepackage{listings}%
\usepackage{mdframed}
\usepackage{adjustbox}
\usepackage{etoolbox}
\usepackage{subcaption}

\theoremstyle{thmstyleone}%
\theoremstyle{thmstyletwo}%

\theoremstyle{thmstylethree}%

\begin{document}

\title[Collisions of Multiple Black Holes: A Numerical Relativity Study]{Collisions of Multiple Black Holes: A Numerical Relativity Study}


\author*[1]{\fnm{Chen-Kai} \sur{Qiao}}\email{chenkaiqiao@cqut.edu.cn}

\author[1]{\fnm{Shu-Yuan} \sur{Liu}} 


\affil[1]{\orgdiv{School of Physics and New Energy}, \orgname{Chongqing University of Technology}, 
\orgaddress{
\city{Chongqing}, \postcode{400054}, 
\country{China}}}


\abstract
{\unboldmath
Numerical relativity has a profound impact on gravitational theory and gravitational wave astronomy. In this work, we present a systematic study of the head-on collisions of multipule black hole systems within the framework of numerical relativity using the Baumgarte-Shapiro-Shibata-Nakamura (BSSN) formulation. The numerical simulations of the dynamical evolution of gravitational fields are computed using the AMSS-NCKU code. We report numerical results for black hole collisions in several symmetric and non-symmetric configurations, in which the black holes are initially at rest and are described by Bowen-York type initial data. Our results suggest that black hole systems with perfectly symmetric configurations (e.g., black holes located at the vertices of a regular polygon or a regular polyhedron) impose strong constraints on the emitted gravitational waves, and no gravitational kick arises in these symmetric systems. For regular polygonal configurations, the radiation is dominated by the plus polarization of the $l=2$, $m=0$ mode. For regular polyhedral configurations, all the quadrupole modes with $l=2$ nearly vanish, and gravitational wave amplitudes are strongly suppressed to the order of $10^{-4}$ or below, which is in accordance with an analysis based on the quadrupole approximation. Gravitational radiations from $m \ne 0$ modes become non‑negligible only when non-symmetric configurations are encountered.  
}




\maketitle

\section{Introduction}\label{sec1}

Numerical relativity plays a central role in theoretical physics and astronomy, enabling precise and first-principle investigations of the dynamical evolution of gravitational fields in the strong-field regime and other extreme physical environments. It has become an indispensable tool for high-precision tests of general relativity through gravitational waves, black hole shadows, and cosmological evolution. In particular, over the past decade, numerical relativity has provided the most accurate predictions for compact binary coalescences and their associated gravitational waves \cite{SXS-Catelog-2013,SXS-Catelog-2019,SXS-Catelog-2025,RIT-Catelog-2017,RIT-Catelog-2019,RIT-Catelog-2020,RIT-Catelog-2022,Maya-Catelog-2016,Maya-Catelog-2025,BAM-Catalog-2024,GWTC-2019}. These waveforms serve as ideal templates for the observational detection and parameter estimation of gravitational wave signals captured by the LIGO-Virgo-KAGRA collaborations, as exemplified by the landmark events GW150914, GW170817, and GW190521 \cite{GW150914-A,GW150914-B,GW170817-A,GW170817-B,GW190521-A,GW190521-B}.

Despite the great success of numerical relativity in modeling binary black hole systems and other compact binary coalescences, its direct application to multiple black hole systems remains far less developed. Nevertheless, multiple black hole systems are of profound astrophysical and theoretical importance. Observationally, these systems can drive hierarchical merger processes, in which the merger remnants of earlier coalescences can subsequently participate in further mergers \cite{Antonini2019,Fragione2023,Kimball2021,LiuB2021,FanYZ2024,HanWB2025}, and the resulting gravitational wave emissions carry unique fingerprints distinct from those of conventional binary black hole systems \cite{Gerosa2021}. Theoretically, they provide examples for examining the nonlinear dynamics of the gravitational field and the intricate behavior of relativistic N‑body systems in highly curved spacetimes \cite{Gultekin2006,Asada2007,Asada2011,Lousto2008a,Lousto2008b,Lousto2023,Lousto2024,Galaviz2010,Galaviz2011,CaoZJ2011,Heinze2025,Heinze2026,Bamber2025,Bonetti2016,Bonetti2018,LiD2021,Doctolero2026}. Unfortunately, simulating the dynamical evolution of general multiple black hole systems (such as the unrestricted relativistic $N$-body problem) in numerical relativity poses a formidable computational challenge. For simplicity, we therefore focus on the head-on collisions of multiple black hole systems that are initially at rest, for which the initial data can be conveniently constructed using the Bowen-York prescription \cite{Bowen-York1980}. 

In Maxwell electrodynamics, it is a well-established result that the symmetry of a charged source imposes stringent constraints on its electromagnetic radiation, governing the selection rules, polarization states, and the multipolar structure of the emitted electromagnetic waves. Gravitational waves, the characteristic wave phenomena of spin-$2$ tensor fields, exhibit a deep analogy with the electromagnetic radiations (wave phenomenons for the spin-$1$ electromagnetic vector fields). It is therefore natural to expect that the symmetry of a gravitating source configuration will exert a significant influence on the emitted gravitational waves, as in the electromagnetic wave cases. Specifically, one can expect that the spatial symmetries can suppress most of the multipolar modes in the gravitational radiation, simplify the waveform morphology, and leave distinct, measurable imprints on gravitational wave signals. Accordingly, a systematic understanding of these symmetry-induced constraints --- on both the dynamical evolution of the black holes and the properties of the resulting gravitational radiations --- is necessary and crucial for general relativity and gravitational wave astronomy.

In this work, we carry out a direct numerical relativity study of multiple black hole systems using the Baumgarte-Shapiro-Shibata-Nakamura (BSSN) formulation \cite{BSSN-1987,BSSN-1995,BSSN-1999}. The BSSN formulation is widely recognized as one of the most successful formulations in numerical relativity, owing to its exceptional numerical stability, its excellent long-term performance for compact-object spacetimes, and its reliable extraction of high-fidelity gravitational waveforms \cite{BS2003,Campanelli2006,BSSN-covariant-2009,BSSN-FO-2012,BSSN-spherical-2012,BSSN-spherical-2013,BSSN-generalized-2015,BSSN-cylindrical-2025,ETK2012,NRPy2018}. Our numerical simulations are carried out using the open-source \texttt{AMSS-NCKU} code, which implements the BSSN formulation to evolve the black hole systems and employs adaptive mesh refinement to ensure high-precision numerical simulations \cite{AMSS-NCKU-2008,AMSS-NCKU-Python}.

The remainder of this paper is organized as follows. In Section \ref{sec2}, we present the theoretical framework and numerical setups. The detailed numerical results for the head-on collisions of multiple black hole systems are presented in Section \ref{sec3}, including the black hole trajectories and the gravitational waveforms. In Section \ref{sec4}, we summarize our findings and conclusions of this work. Throughout this work, geometrized units with $G=c=1$ are adopted.

\section{Theoretical Framework and Numerical Setup}\label{sec2}

In this section, we summarize the theoretical framework and numerical setup used in this work. In our work, the dynamical evolution of multiple black hole systems is carried out using the Baumgarte-Shapiro-Shibata-Nakamura (BSSN) formulation in numerical relativity.

In general relativity, the variations of gravitational fields and the motions of black holes are governed by Einstein's gravitational field equations
\begin{equation}
	R_{\mu\nu} - \frac{1}{2} R g_{\mu\nu} = 8 \pi T_{\mu\nu} . \label{Einstein Equation}
\end{equation}
The Einstein field equations form a set of highly nonlinear partial differential equations, for which analytical solutions are difficult to obtain, and they are usually solved numerically with high-performance simulation codes. To solve these equations numerically, it is necessary to employ the $3+1$ decomposition of the 4-dimensional Lorentzian manifold, which enables us to recast Einstein's equations into a Cauchy initial-value problem. Under the conventional Arnowitt-Deser-Misner (ADM) $3+1$ decomposition, the spacetime metric takes the form
\begin{equation}
	ds^{2} = - \alpha^{2} dt^{2} + \gamma_{ij} ( dx^{i} + \beta^{i} dt ) ( dx^{j} + \beta^{j} dt ) ,
\end{equation}
where $\gamma_{ij}$ is the induced metric on the 3-dimensional spacelike hypersurface, and $\alpha$, $\beta^{i}$ are the lapse function and shift vector, respectively. Once appropriate initial data at $t=t_{0}$ are specified, the subsequent evolution of the gravitational field and spatial metric $\gamma_{ij}$ can be obtained by solving a set of numerically solvable partial differential equations relevant to the time coordinate $t$.
Furthermore, in the reduction of Einstein's gravitational field equation (\ref{Einstein Equation}) into a numerically stable first-order evolution system, different choices of the auxiliary variable and gauge variables lead to different kinds of differential formulations, such as Baumgarte-Shapiro-Shibata-Nakamura (BSSN) formulation \cite{BSSN-1987,BSSN-1995,BSSN-1999}, conformal and covariant Z4 (CCZ4) formulation \cite{Z4-2003,Z4-2004,CCZ4-2012,CCZ4-2013,CCZ4-2018}, and the conformal Z4 formulation with constraint damping (Z4c) \cite{Z4C-2010,Z4C-2012,Z4C-2013}.

In this work, the Baumgarte-Shapiro-Shibata-Nakamura (BSSN) equations are implemented to carry out the numerical simulations. The BSSN equations, proposed by T. W. Baumgarte, S. L. Shapiro, M. Shibata, T. Nakamura \cite{BSSN-1987,BSSN-1995,BSSN-1999} \footnote{Some literature also named the Baumgarte-Shapiro-Shibata-Nakamura (BSSN) equations as Baumgarte-Shapiro-Shibata-Nakamura-Oohara-Kojima (BSSNOK) equations.}, have become the standard framework for simulating the dynamical evolution of black hole systems and neutron star systems. To write the BSSN equations explicitly, it is necessary to decompose the spatial metric $\gamma_{ij}$ into conformal metric $\bar{\gamma}_{ij}$ with unit determinant and a conformal factor $\phi$
\begin{equation}
	\gamma_{ij} = e^{4\phi} \tilde{\gamma}_{ij} .
\end{equation}
In addition, it is also needed to decompose the extrinsic curvature of hypersurface $K_{ij}$ into its trace part $\text{tr}K = \gamma^{ij} K_{ij}$ and a traceless part $A_{ij}$, via
\begin{equation}
	K_{ij} = \underbrace{ A_{ij} }_{\text{traceless part}} + \underbrace{ \frac{1}{3} \gamma_{ij} \cdot \text{tr}K }_{\text{trace part}}
	= e^{4\phi} \tilde{A}_{ij} + \frac{1}{3} \gamma_{ij} \cdot \text{tr}K ,
\end{equation}
where $\tilde{A}_{ij} \equiv e^{-4\phi} A_{ij}$ is the conformally rescaled traceless extrinsic curvature. Substituting these decomposed variables into Einstein's gravitational field equations and performing the standard $3+1$ reduction, one eventually obtains the computational equation form required for numerical relativity simulations (the BSSN evolution equations) 
\begin{subequations}
	\begin{eqnarray}
		\partial_{t} \phi
		& = & - \frac{1}{6} \alpha \cdot \text{tr}K + \frac{1}{6} \cdot \partial_{k} \beta^{k}
		+ \beta^{i} \cdot \partial_{i} \phi , 
		\label{BSSN equation phi}
		\\
		\partial_{t} \tilde{\gamma}_{ij} 
		& = & - 2 \alpha \tilde{A}_{ij} 
		+ \beta^{k} \cdot \partial_{k} \tilde{\gamma}_{ij} 
		+ \tilde{\gamma}_{ik} \cdot \partial_{j} \beta^{k}
		+ \tilde{\gamma}_{kj} \cdot \partial_{i} \beta^{k}
		- \frac{2}{3} \tilde{\gamma}_{ij} \cdot \partial_{k} \beta^{k} ,
		\label{BSSN equation gamma}
		\\
		\partial_{t} \text{tr}K
		& = & \alpha \bigg[ \tilde{A}_{ij} \tilde{A}^{ij} 
		+ \frac{1}{3} (\text{tr}K)^{2} \bigg]
		- \gamma^{ij} D_{i} D_{j} \alpha 
		+ 4 \pi \alpha (\rho + S)
		+ \beta^{i} \cdot \partial_{i} (\text{tr}K) ,
		\label{BSSN equation K}
		\\
		\partial_{t} \tilde{A}_{ij}
		& = & e^{-4\phi} 
		\bigg[ \alpha R_{ij} - D_{i} D_{j} \alpha - 8 \pi \alpha S_{ij} \bigg]^{\text{TF}}
		+ \alpha \bigg[ \text{tr}K \cdot \tilde{A}_{ij} - 2 \tilde{A}_{ik} \tilde{A}_{j}^{k} \bigg]  
		\nonumber
		\\
		&   & + \beta^{k} \cdot \partial_{k} \tilde{A}_{ij}
		+ \tilde{A}_{ik} \cdot \partial_{j} \beta^{k}
		+ \tilde{A}_{kj} \cdot \partial_{i} \beta^{k}
		- \frac{2}{3} \tilde{A}_{ij} \cdot \partial_{k} \beta^{k} ,
		\label{BSSN equation A}
		\\
		\partial_{t} \tilde{\Gamma}^{i}
		& = & - 2 \alpha \tilde{A}^{ik} \cdot \partial_{k} \alpha
		+ 2 \alpha \bigg[ \tilde{\Gamma}^{i}_{jk} \tilde{A}^{jk} 
		- \frac{2}{3} \tilde{\gamma}^{ik} \cdot \partial_{k} (\text{tr}K) 
		- 8 \pi \tilde{\gamma}^{ik} S_{k} 
		+ 6 \tilde{A}^{ik} \cdot \partial_{k} \phi \bigg] 
		\nonumber
		\\
		&   & + \beta^{k} \cdot \partial_{k} \tilde{\Gamma}^{i}
		- \tilde{\Gamma}^{k} \cdot \partial_{k} \beta^{i}
		+ \frac{2}{3} \tilde{\Gamma}^{i} \cdot \partial_{k} \beta^{k}
		+ \frac{1}{3} \tilde{\gamma}^{ik} \cdot \partial_{k} \partial_{j} \beta^{j}
		+ \tilde{\gamma}^{kj} \cdot \partial_{j} \partial_{k} \beta^{i} .
		\label{BSSN equation Lambda}
	\end{eqnarray}
\end{subequations}
In these equations, the superscript $[...]^{\text{TF}}$ denotes the trace-free part (traceless part) of terms in the brackets, $D_{i}$ represents the covariant derivative compatible with the physically 3 dimensional metric $\gamma_{ij}$, $R_{ij}$ is Ricci curvature of the spatial metric $\gamma_{ij}$, and $\rho$, $S_k$, $S_{ij}$ denote the components from the energy-momentum-stress tensor (which corresponds to the energy density, momentum density, and spatial stress). In this formulation, the set of variables  $\{\phi, \tilde{\gamma}_{ij}, \text{tr}K, \tilde{A}_{ij}, \tilde{\Gamma}^{i} \}$ are served as the dynamical variables governing the time evolution of gravitational fields and spacetime structures. It is also worth noting that these dynamical variables are subject to several constraints. In other words, in addition to the BSSN equations (\ref{BSSN equation phi})-(\ref{BSSN equation Lambda}), the dynamical variables also satisfy the following Hamiltonian and momentum constraint equations
\begin{subequations}
\begin{eqnarray}
	\text{Hamiltonian Constraint} 
	& \Leftrightarrow & 
	\mathcal{H} = R^{(3)} - K_{ij} K^{ij} + (\text{tr}K)^2 - 16 \pi \rho = 0 ,
	\\
	\text{Momentum Constraint} 
	& \Leftrightarrow & 
	\mathcal{M}^{i} = D_{j} ( K^{ij} - \gamma^{ij} \text{tr}K ) - 8 \pi S^{i} = 0 ,
\end{eqnarray}
\end{subequations}
where $R^{(3)}$ is the scalar curvature of the spatial metric. By numerically solving the dynamical BSSN equations and the constraint equations on high-resolution computational grids (usually with adaptive mesh refinement techniques), we are able to perform numerical relativity simulations to calculate the evolution of the gravitational fields and the trajectories of compact celestial bodies.

For black holes that are initially placed at rest, the corresponding initial data can be constructed using the Bowen-York prescription \cite{Bowen-York1980}, in which the initial data set $\{\phi, \tilde{\gamma}_{ij}, \text{tr}K, \tilde{A}_{ij}, \tilde{\Gamma}^{i} \}$ is given by
\begin{subequations}
\begin{eqnarray}
	\psi &\equiv& e^{\phi} = 1 + \sum_{I=1}^{N} \frac{m_{I}}{2|\mathbf{r}-\mathbf{r_{I}}|} ,
	\\
	\tilde{\gamma}_{ij} &=& \delta_{ij}, \ \ 
	\tilde{A}_{ij} = 0, \ \ 
	\text{tr}K = 0, \ \
	\tilde{\Gamma}^{i} = 0 ,
\end{eqnarray}
\end{subequations}
with $m_{I}$ to be the bare mass of each black hole, and $|\mathbf{r}-\mathbf{r_{I}}|$ denotes the modulus of the position vector $\mathbf{r}-\mathbf{r_{I}}$), and $\delta_{ij}$ is the Kronecker delta that denotes the flat reference metric.

In the present work, we use the \texttt{AMSS-NCKU} code to perform the numerical relativity simulations of the multiple black hole collisions. \texttt{AMSS-NCKU} is an open-source numerical general relativity code developed by a Chinese research group \cite{AMSS-NCKU-2008}. This code specializes in high-precision simulations of binary and multiple black hole systems and features a self-contained parallel implementation of adaptive mesh refinement (AMR). It supports the black hole numerical simulations with various formulations of computational equations, including the BSSN equations \cite{AMSS-NCKU-2008,AMSS-NCKU-BSSN-2012,AMSS-NCKU-BSSN-2015}, Z4c equations \cite{Z4C-2012,Z4C-2013}, BSSN equations coupled with electromagnetic fields, and BSSN equations coupled with scalar fields in $f(R)$ gravity \cite{AMSS-NCKU-FR}. 
Furthermore, an open-source and user-friendly Python interface for the \texttt{AMSS-NCKU} code has recently been developed \cite{AMSS-NCKU-Python}, which enables the automatic initialization and execution of the simulations, as well as the automatic visualization of the output data, significantly simplifying the workflow for performing numerical relativity simulations using \texttt{AMSS-NCKU}. 
In practical simulations, the gravitational waveforms are extracted from the \texttt{AMSS-NCKU} outputs by evaluating the multipole expansion of the Newman‑Penrose Weyl scalar $\Psi_{4}$
\begin{subequations}
\begin{eqnarray}		
	&&
	\Psi_{4}(t,r,\theta,\phi) = \sum_{l=2}^{\infty} \sum_{m=-l}^{l} \Psi_{4}^{lm} (t,r) \cdot Y_{s=-2}^{lm}(\theta,\phi) ,
	\\
	&&
	h(t,r,\theta,\phi) = \sum_{l=2}^{\infty} \sum_{m=-l}^{l} h_{lm}(t,r) \cdot Y_{s=-2}^{lm}(\theta,\phi) ,
	\\
	&&
	h_{lm} (t,r) = h^{+}_{lm} (t,r) + i h^{\times}_{lm} (t,r)
	= \int_{-\infty}^{t} dt' \int_{-\infty}^{t'} dt'' \Psi_{4}^{lm} (t'',r) ,
\end{eqnarray}
\end{subequations}
where $h_{+}$ and $h_{\times}$ are the two gravitational wave polarizations in the transverse-traceless gauge, $\Psi_{4}^{lm}$ are the multipole components of the Weyl scalar, and $Y_{s=-2}^{lm}$ denote the spin-weighted spherical harmonics with spin weight $s=-2$ \footnote{In the far-field region, the multipole coefficients $\Psi_{4}^{lm}(t,r)$, $h_{+}^{lm}(t,r)$, $h_{\times}^{lm}(t,r)$ all fall off as $\frac{1}{r}$ with the extraction radius.}.

In the numerical simulation, we adopt the box-to-box AMR structure in \texttt{AMSS-NCKU} (which ensures $\Delta h_{l} = \frac{1}{2} \Delta h_{l-1}$ for two adjacent refinement levels), with 6 basic static grid levels and 4 additional moving grid levels. The grid points for each static box and moving box are $128 \times 128 \times 128$ and $72\times72\times72$ (some simulations adopt $56\times56\times56$ grid points for moving boxes for systems with a larger black hole number). The largest box extents to $-400M \le x,y,z \le +400M$, such that the highest grid resolution is roughly $\Delta h_{\text{highest}} \approx 0.0122M$. In addition, in the \texttt{AMSS-NCKU} code, the spacetime is evolved in the moving puncture gauge, namely the $1+\log$ slicing condition for the lapse $\alpha$ and the Gamma-driver condition for the shift vector $\beta^{i}$ are used \cite{AMSS-NCKU-2008}.

\section{Numerical Results for Collision of Multiple Black Holes}\label{sec3}

In this section, we present the numerical results for the collision of black hole systems from the \texttt{AMSS-NCKU} simulations. We begin with black hole systems exhibiting perfectly symmetric configurations, such as regular polygonal or polyhedral arrangements (where black holes of equal mass $m_{I}=\frac{M}{N}$ are located at the vertices of a regular polygon or a regular polyhedron \footnote{Here, $M=\sum_{I=1}^{N}$ denotes the total mass of the black hole system.}). After investigating these symmetric configurations, we turn to the black hole systems with non-symmetric configurations. Comparisons between the symmetric and asymmetric cases allow us to better examine the impact of spatial symmetry on gravitational wave emissions.

\begin{figure*}
	\centering
	\includegraphics[width=0.425\textwidth]{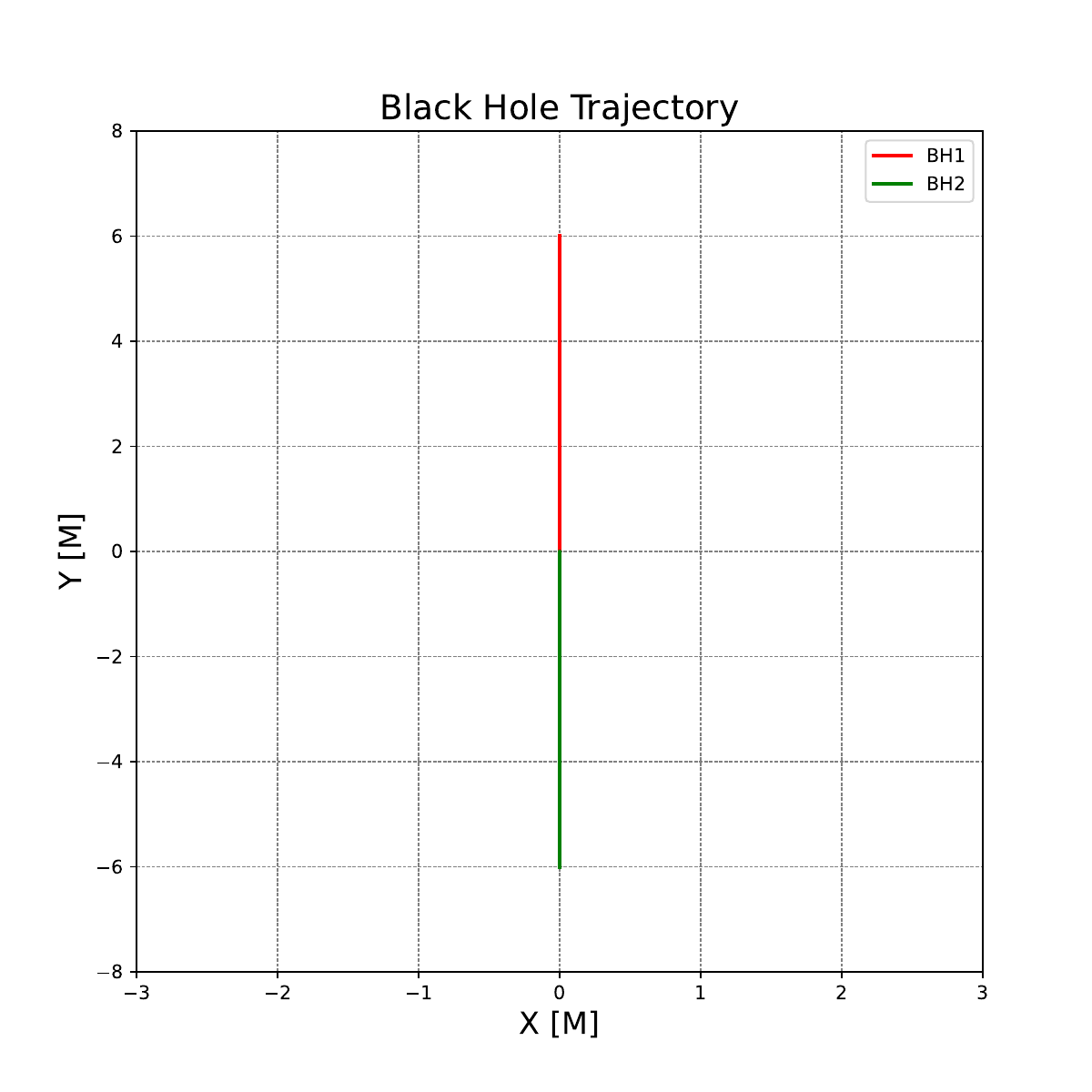}
	\includegraphics[width=0.425\textwidth]{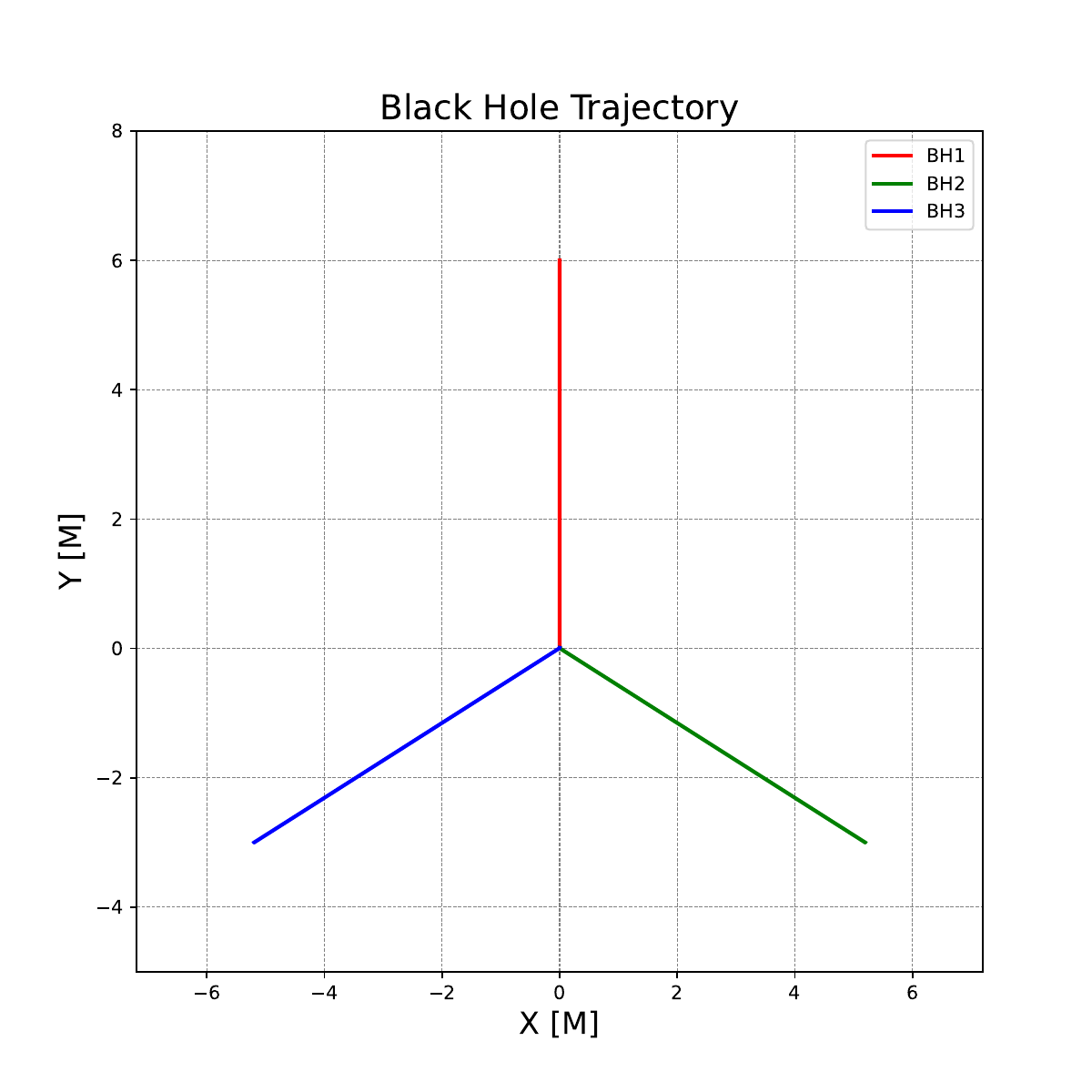}
	\includegraphics[width=0.425\textwidth]{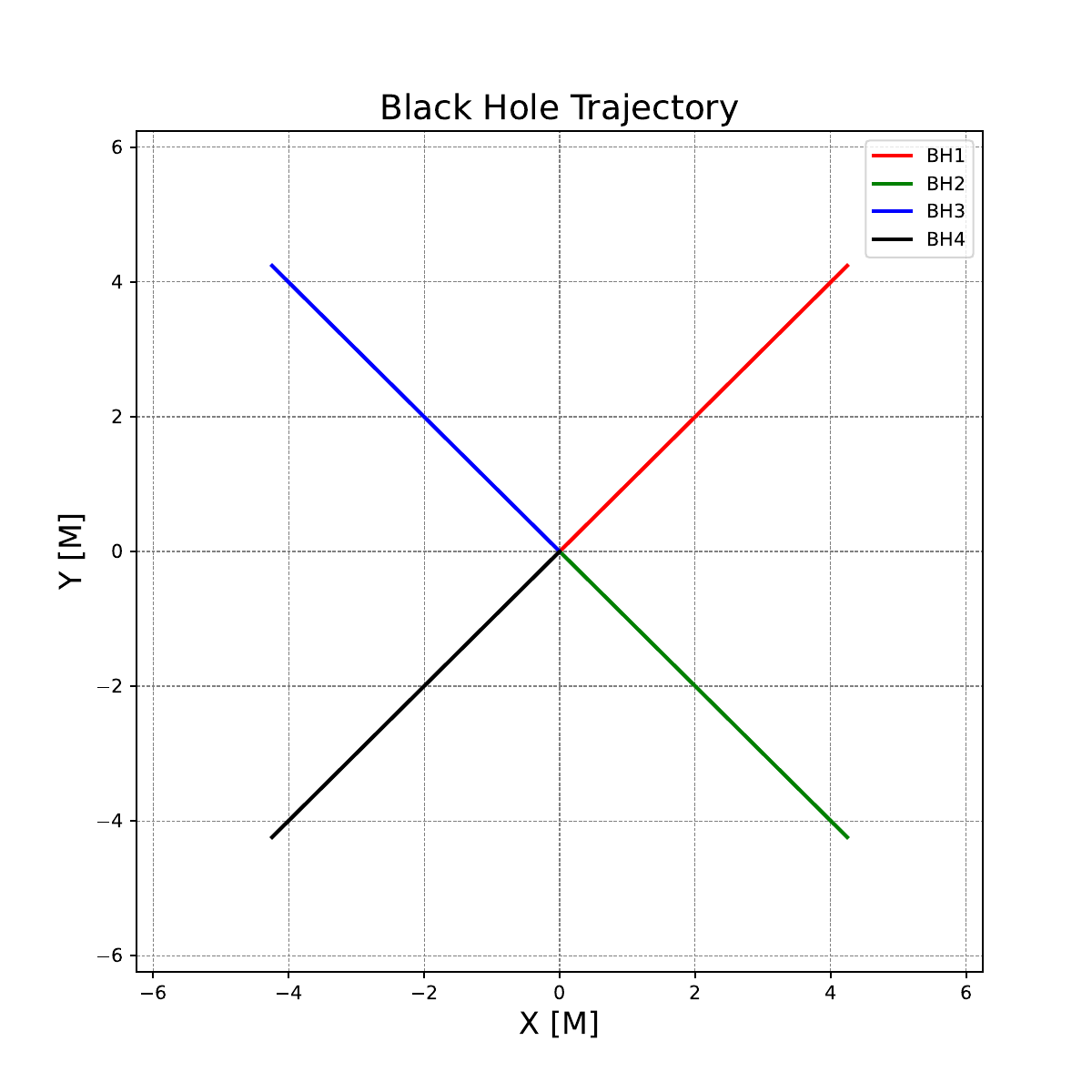}
	\includegraphics[width=0.425\textwidth]{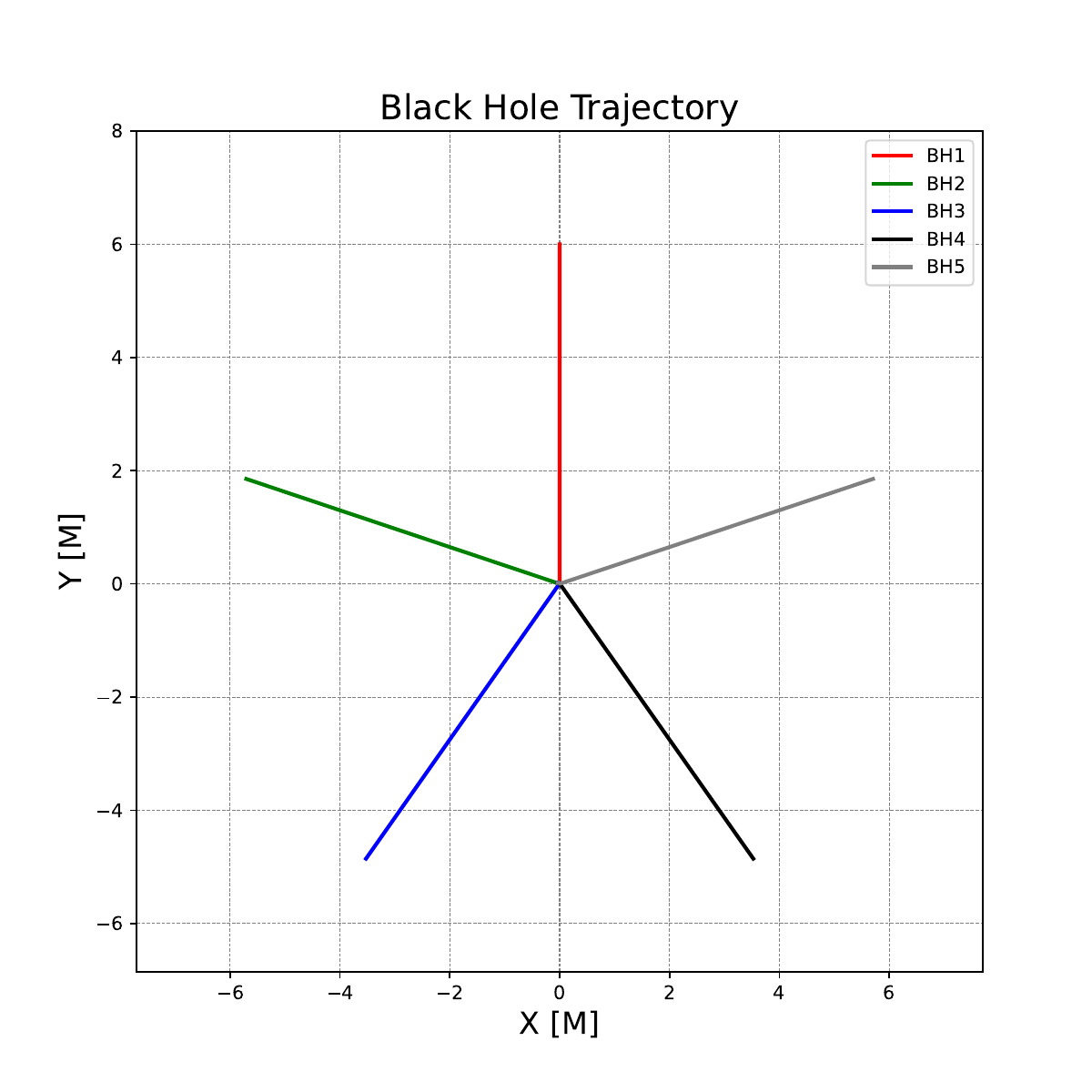}
	\includegraphics[width=0.425\textwidth]{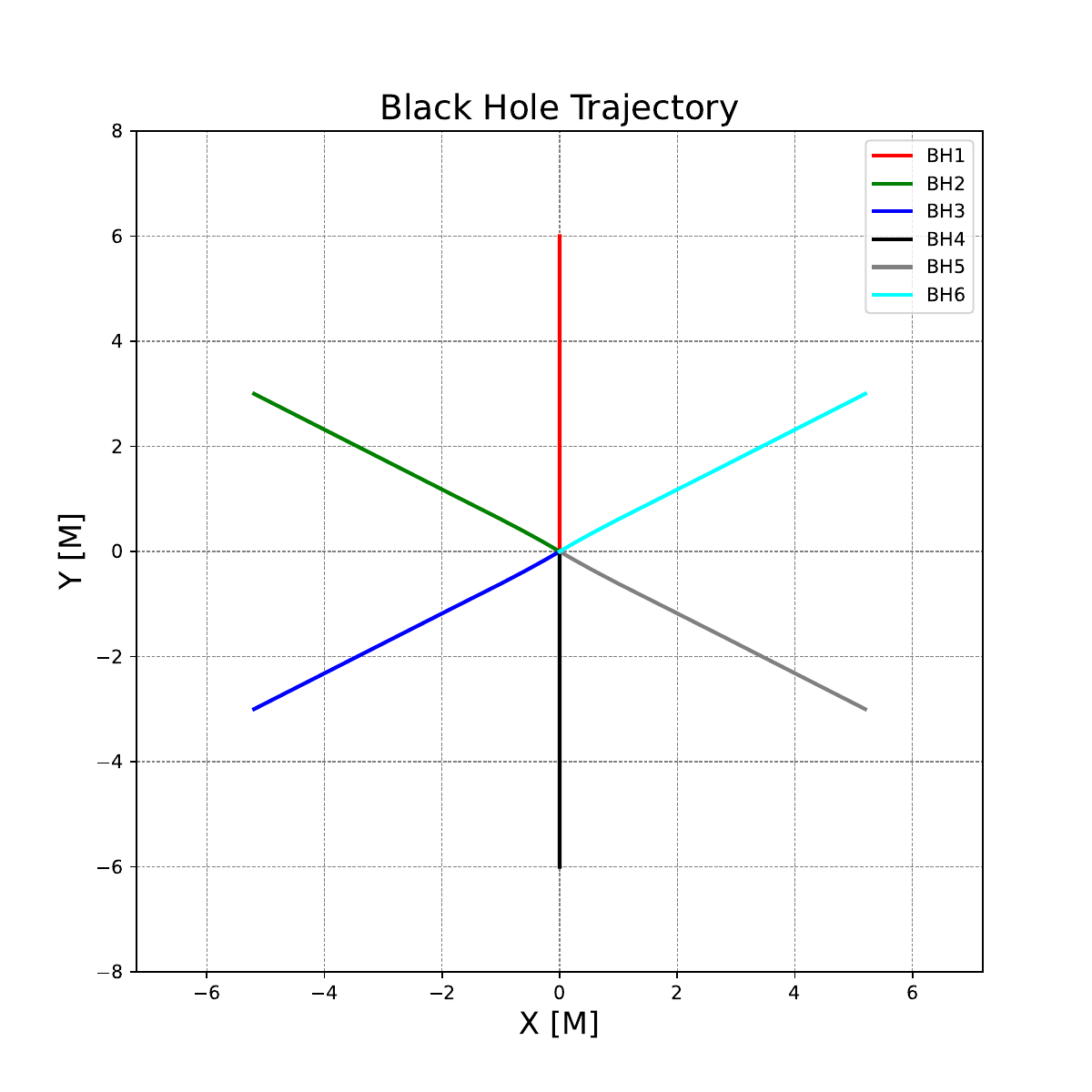}
	\includegraphics[width=0.425\textwidth]{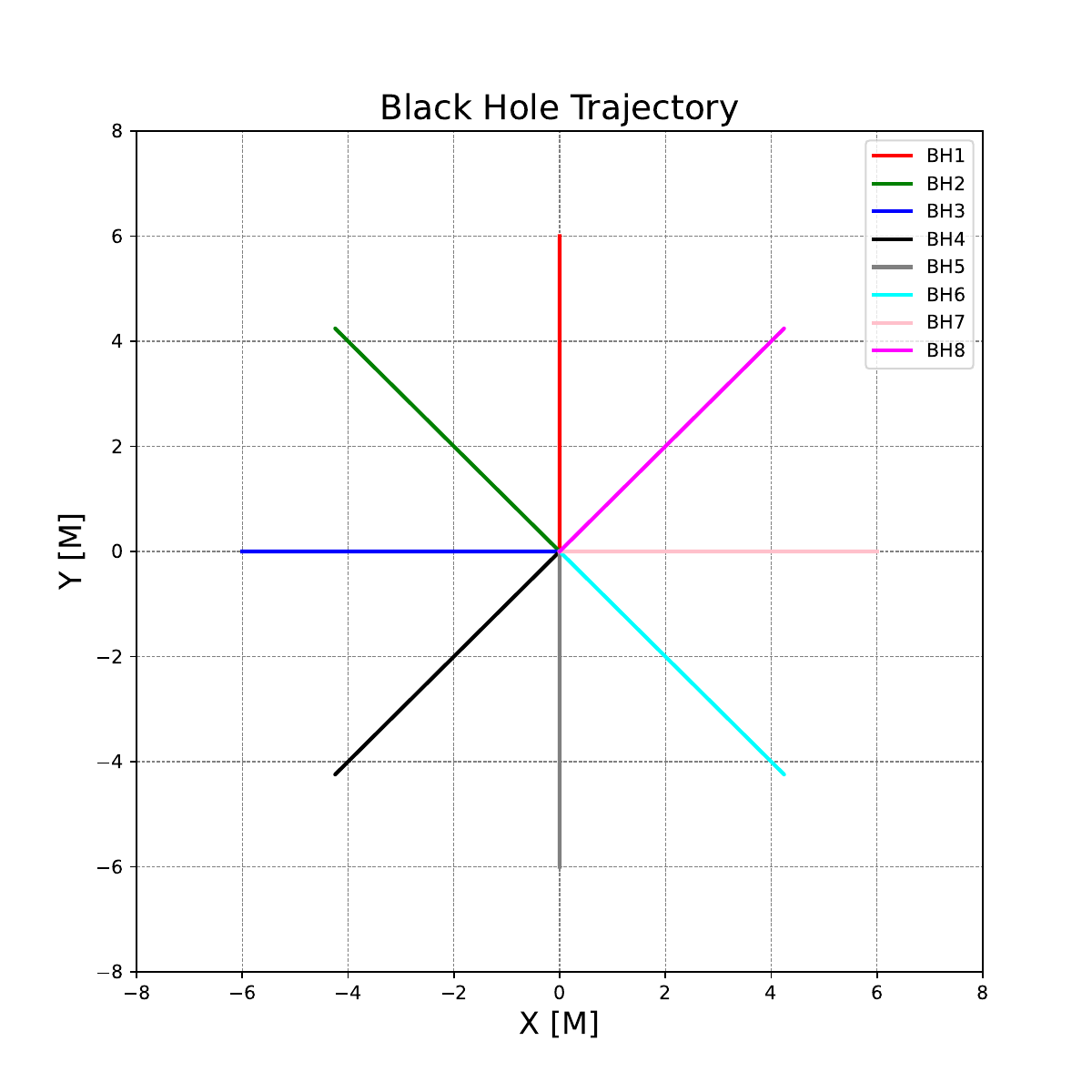}
	\caption{Trajectories of black holes in the $xy$ plane during the head-on collision process, for regular polygonal configurations with $N = 2, 3, 4, 5, 6$, and $8$.}
	\label{Fig BH Trajectory}
\end{figure*}

\begin{figure*}
	\centering
	\includegraphics[width=0.425\textwidth]{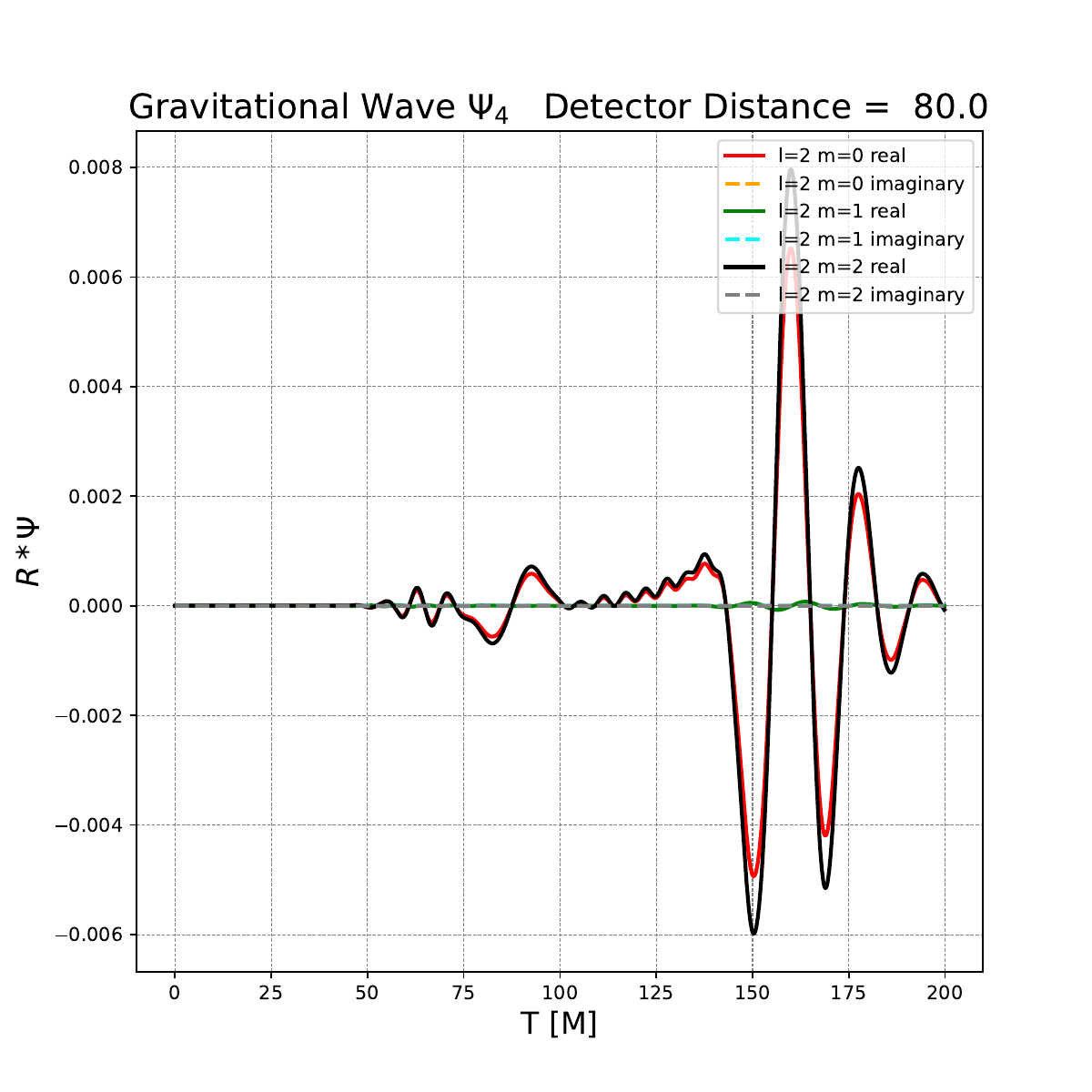}
	\includegraphics[width=0.425\textwidth]{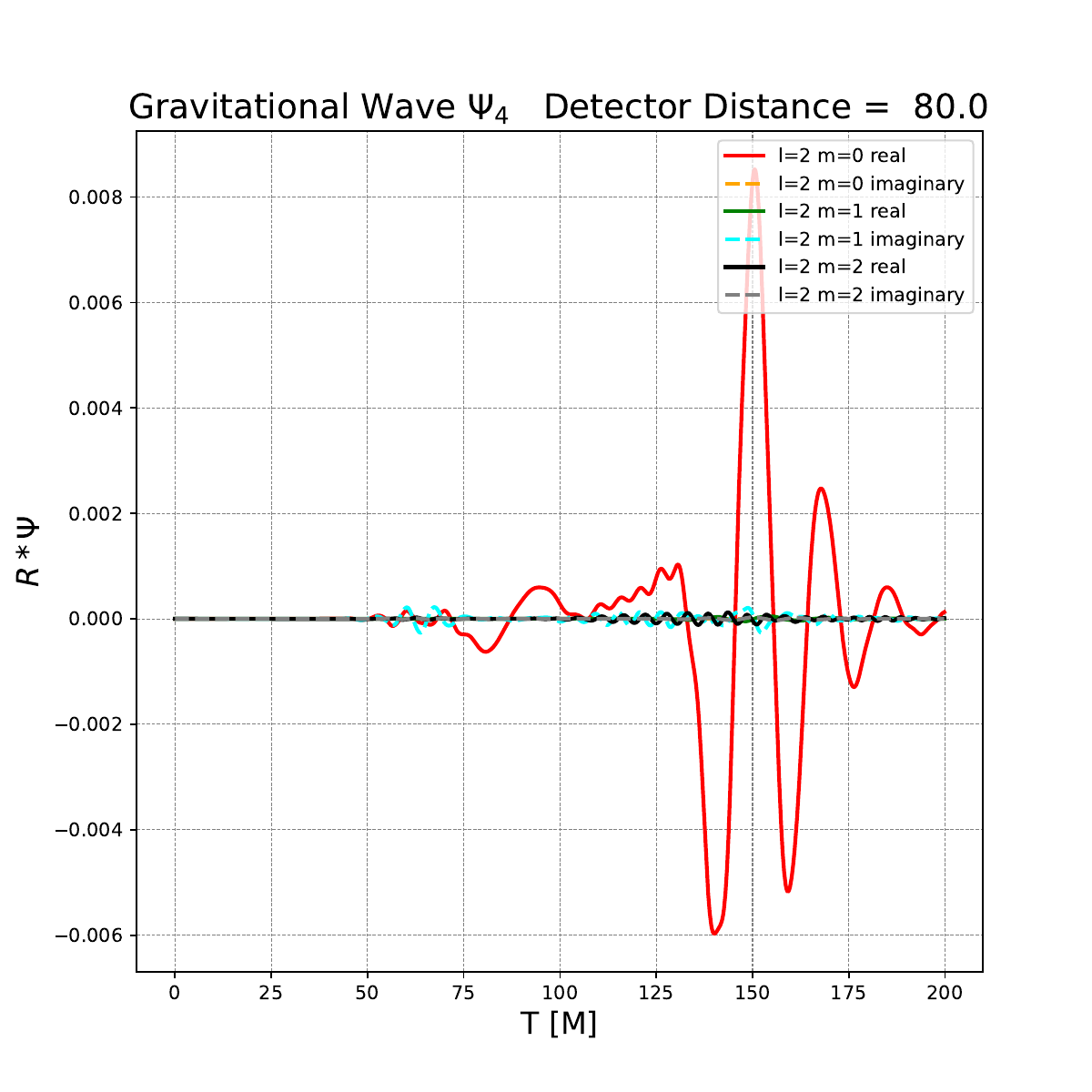}
	\includegraphics[width=0.425\textwidth]{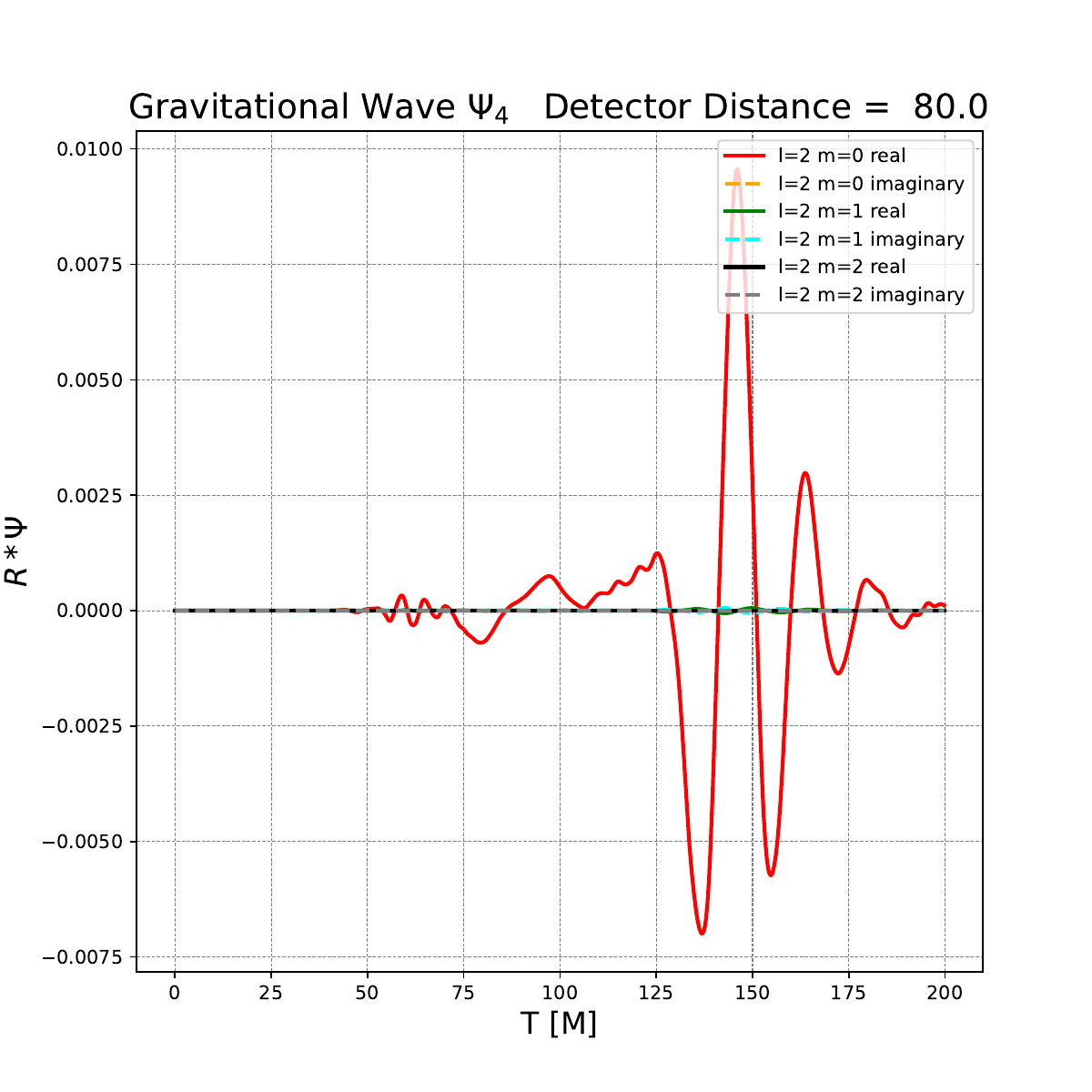}
	\includegraphics[width=0.425\textwidth]{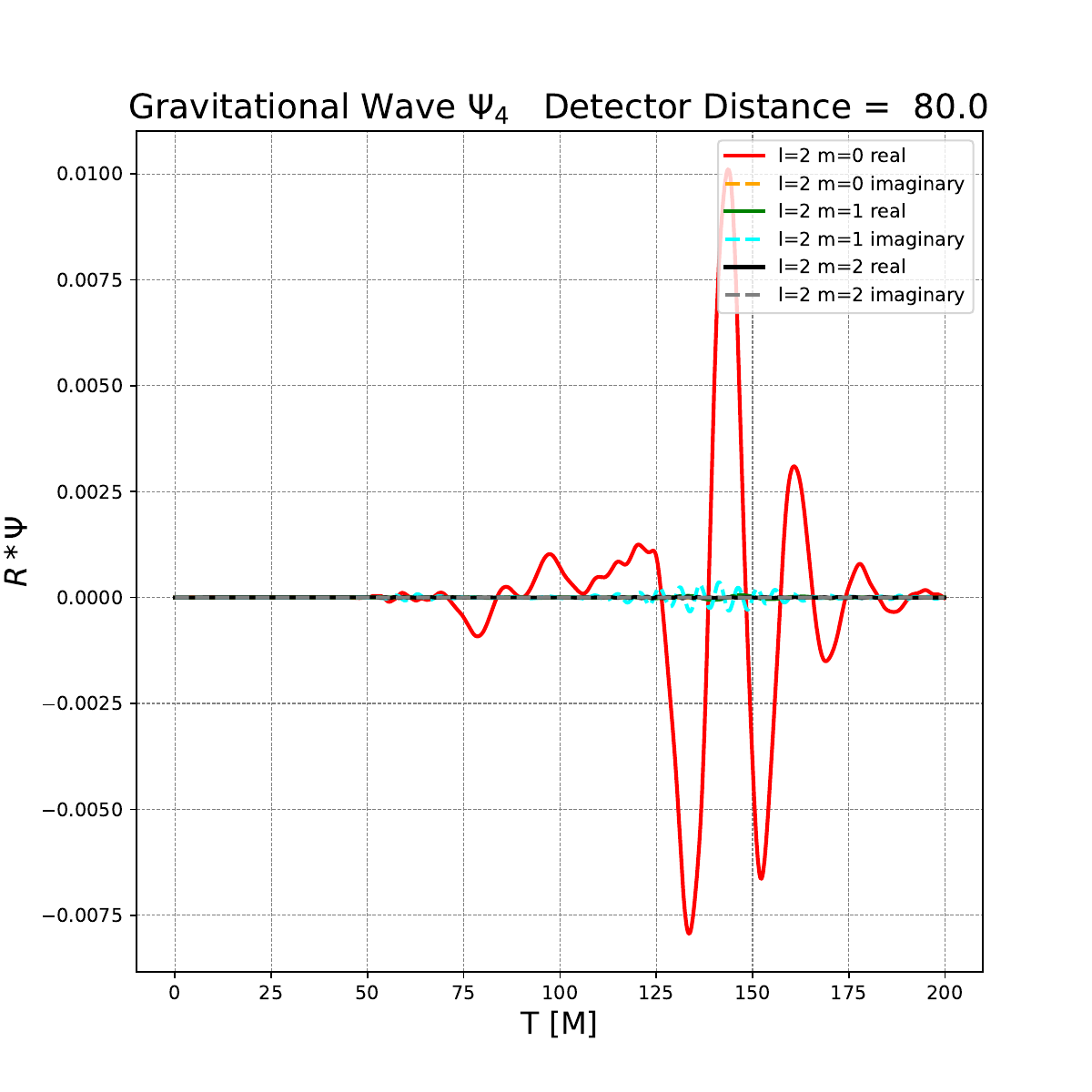}
	\includegraphics[width=0.425\textwidth]{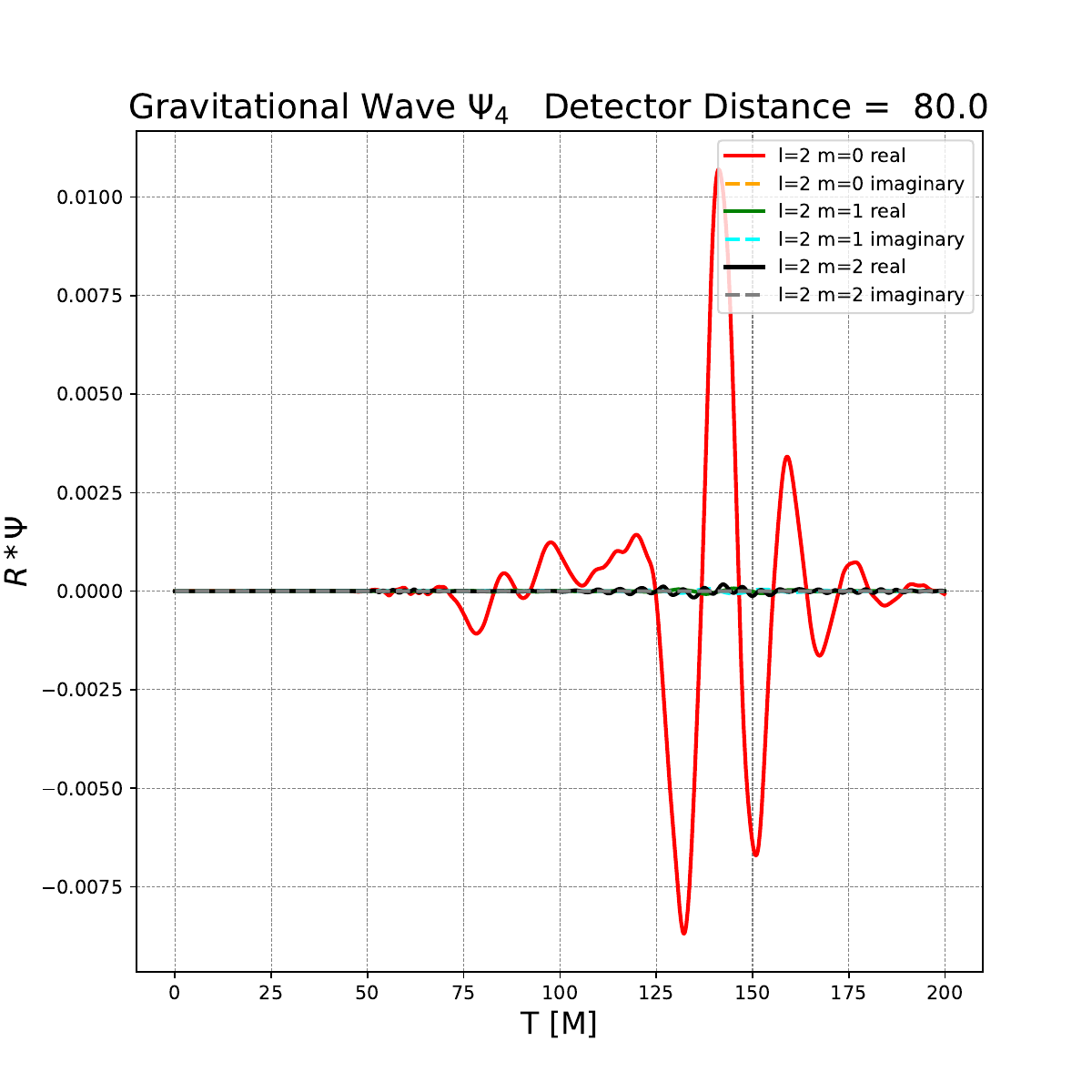}
	\includegraphics[width=0.425\textwidth]{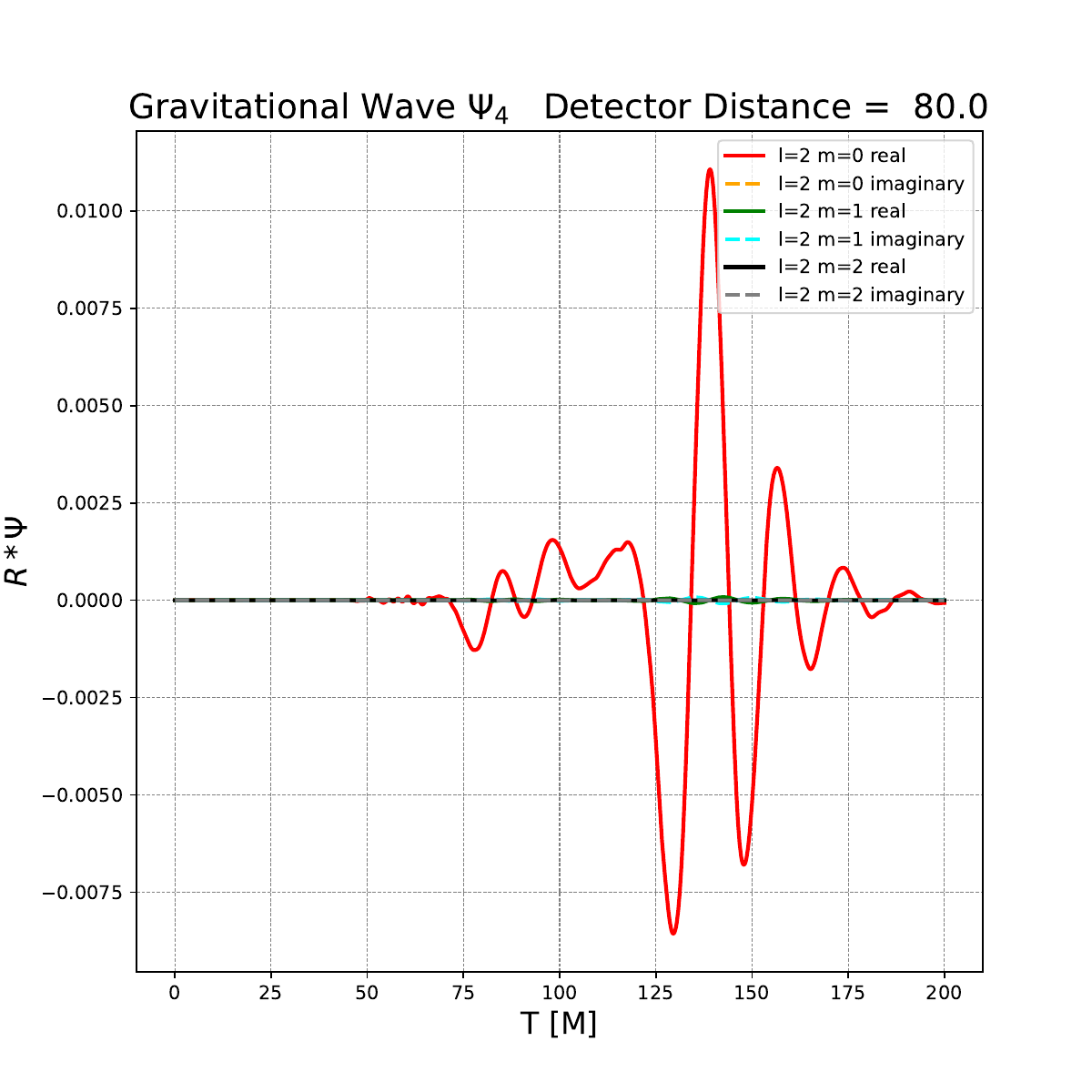}
	\caption{The gravitational waveform of the Weyl scalar $\Psi_4$ in the Newman‑Penrose formalism for multiple black hole systems during the collision process. In this figure, the vertical axis labels the product $r \cdot \Psi_4$, in which the waveform $\Psi_4$ is extracted at the radius $r=80M$. The horizontal axis labels the physical evolution time of the black hole systems in the geometric unit.}
	\label{Fig BH Psi4}
\end{figure*}

\begin{figure*}
	\centering
	\includegraphics[width=0.425\textwidth]{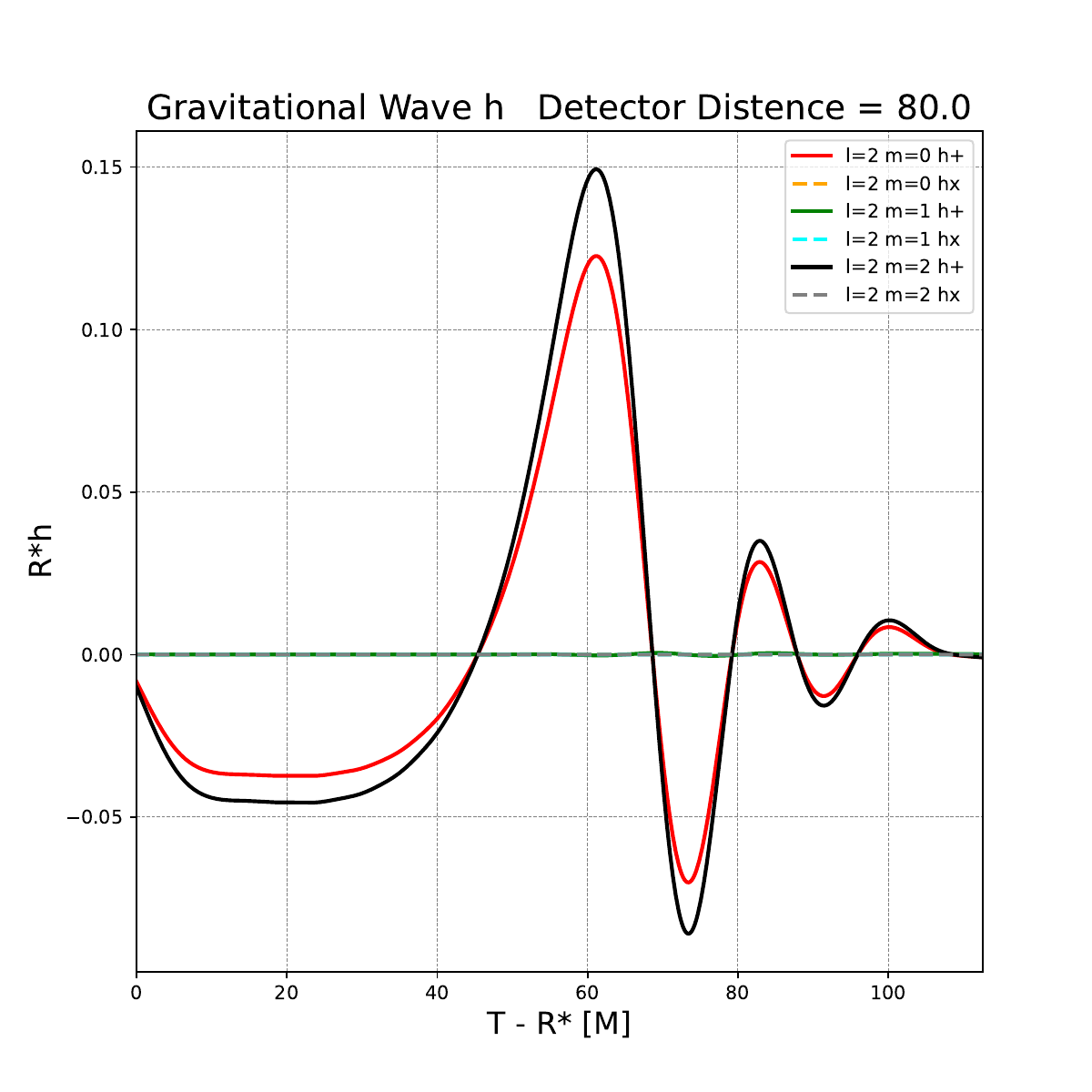}
	\includegraphics[width=0.425\textwidth]{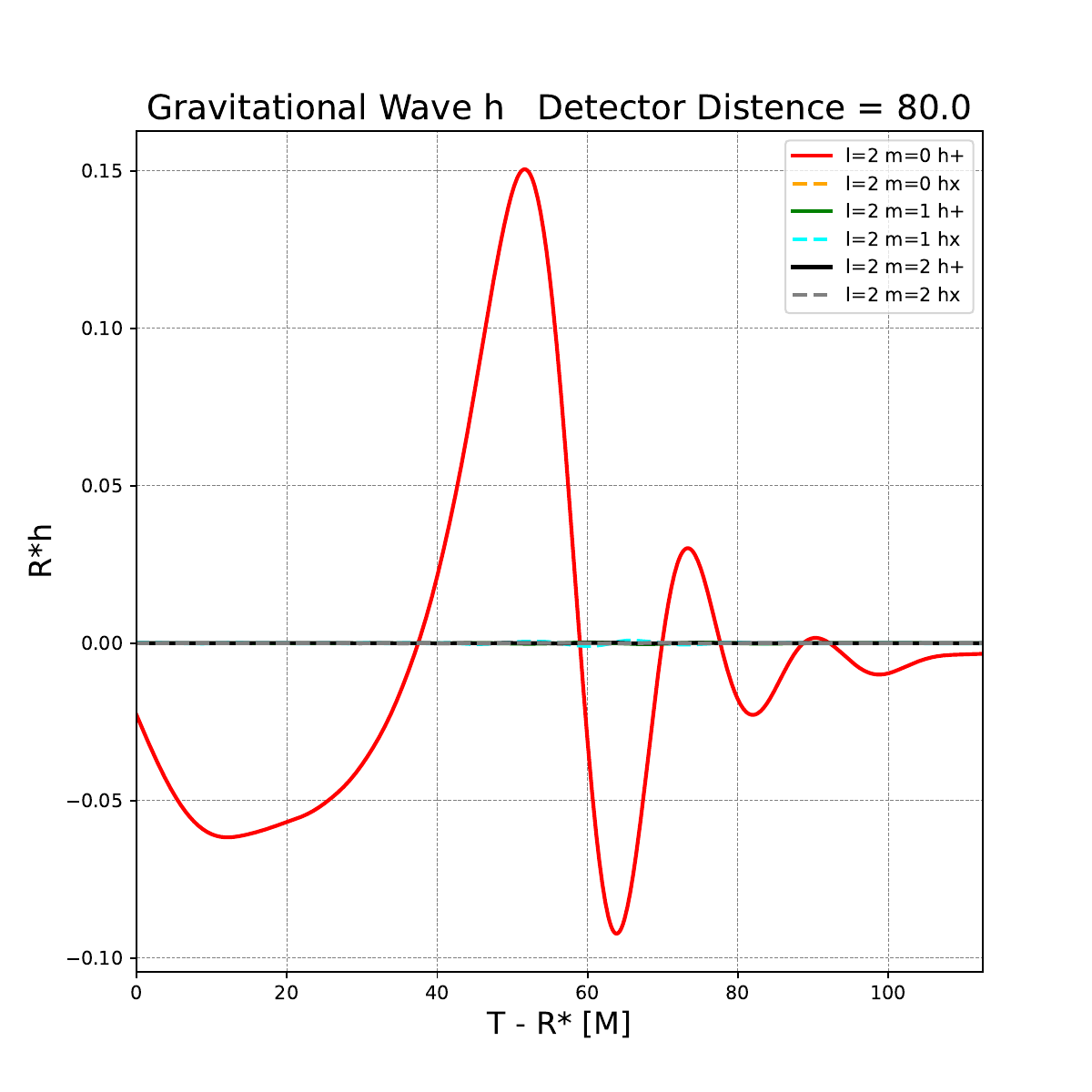}
	\includegraphics[width=0.425\textwidth]{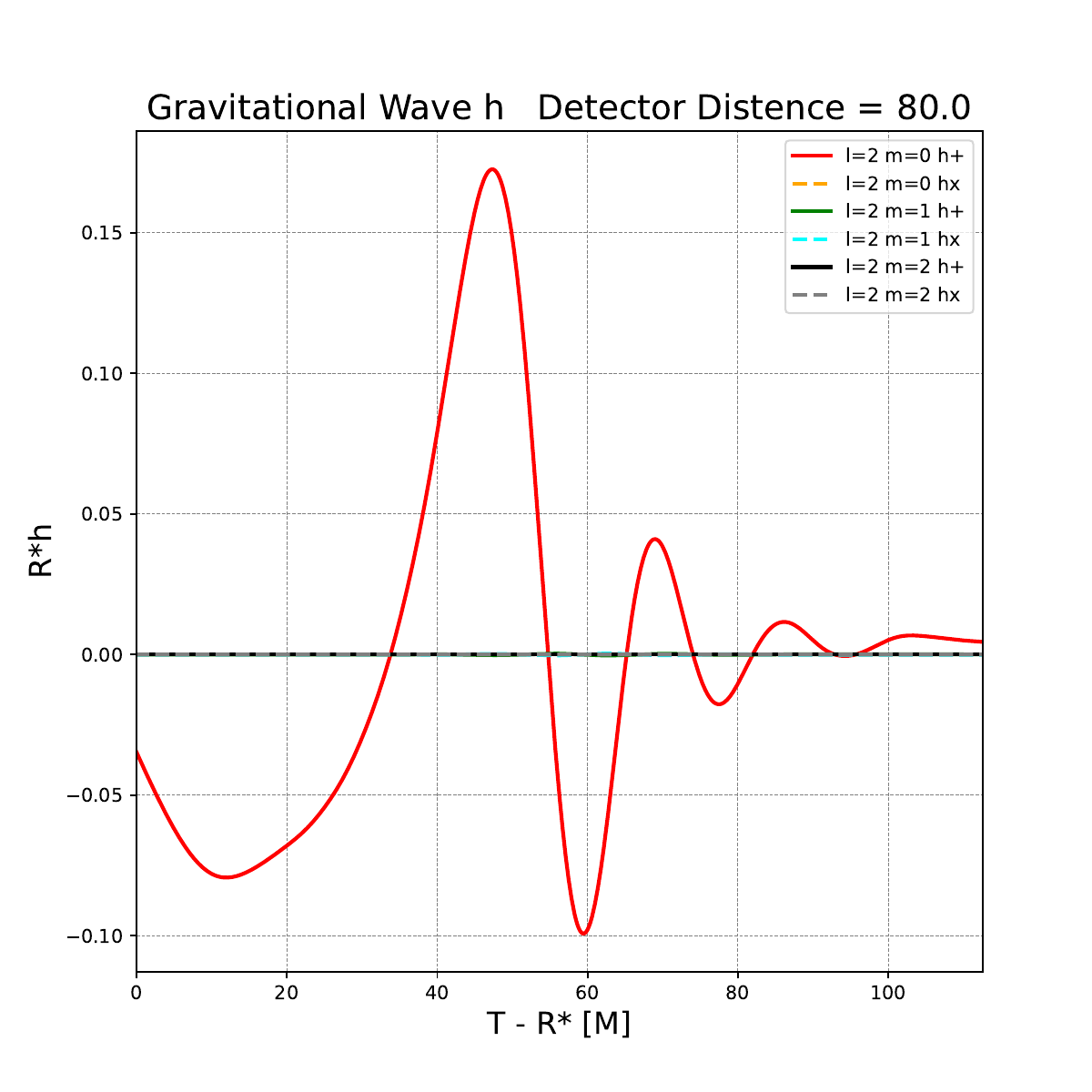}
	\includegraphics[width=0.425\textwidth]{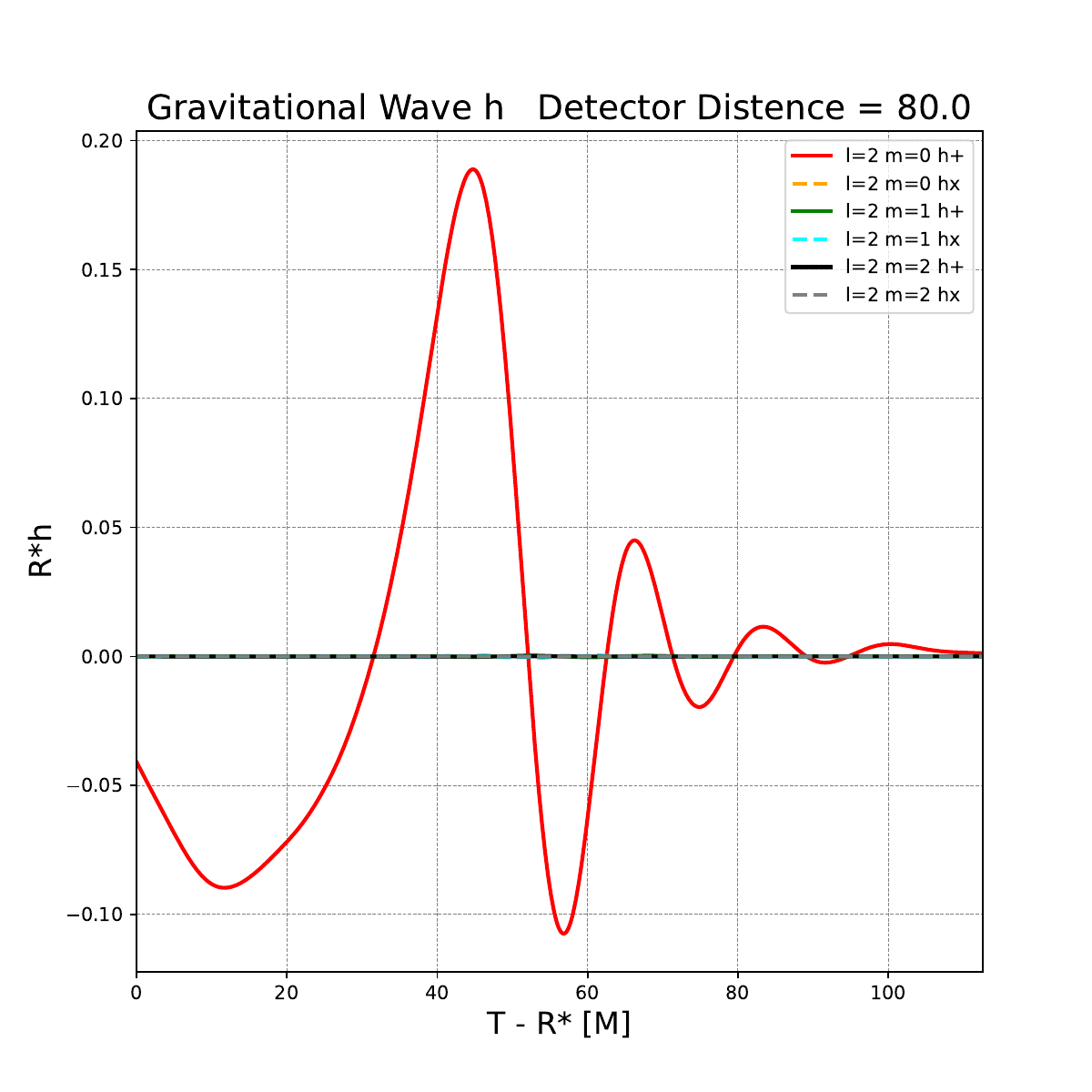}
	\includegraphics[width=0.425\textwidth]{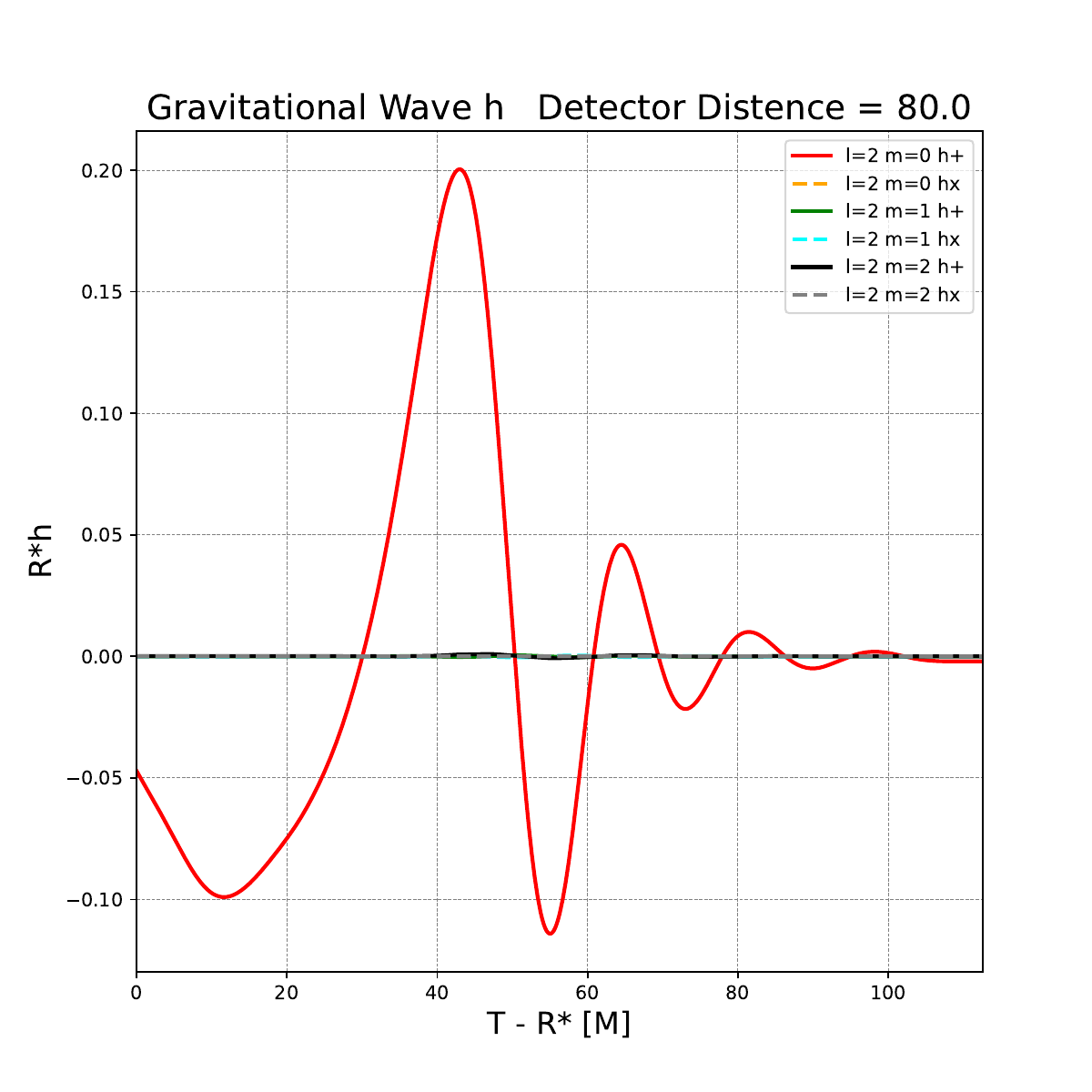}
	\includegraphics[width=0.425\textwidth]{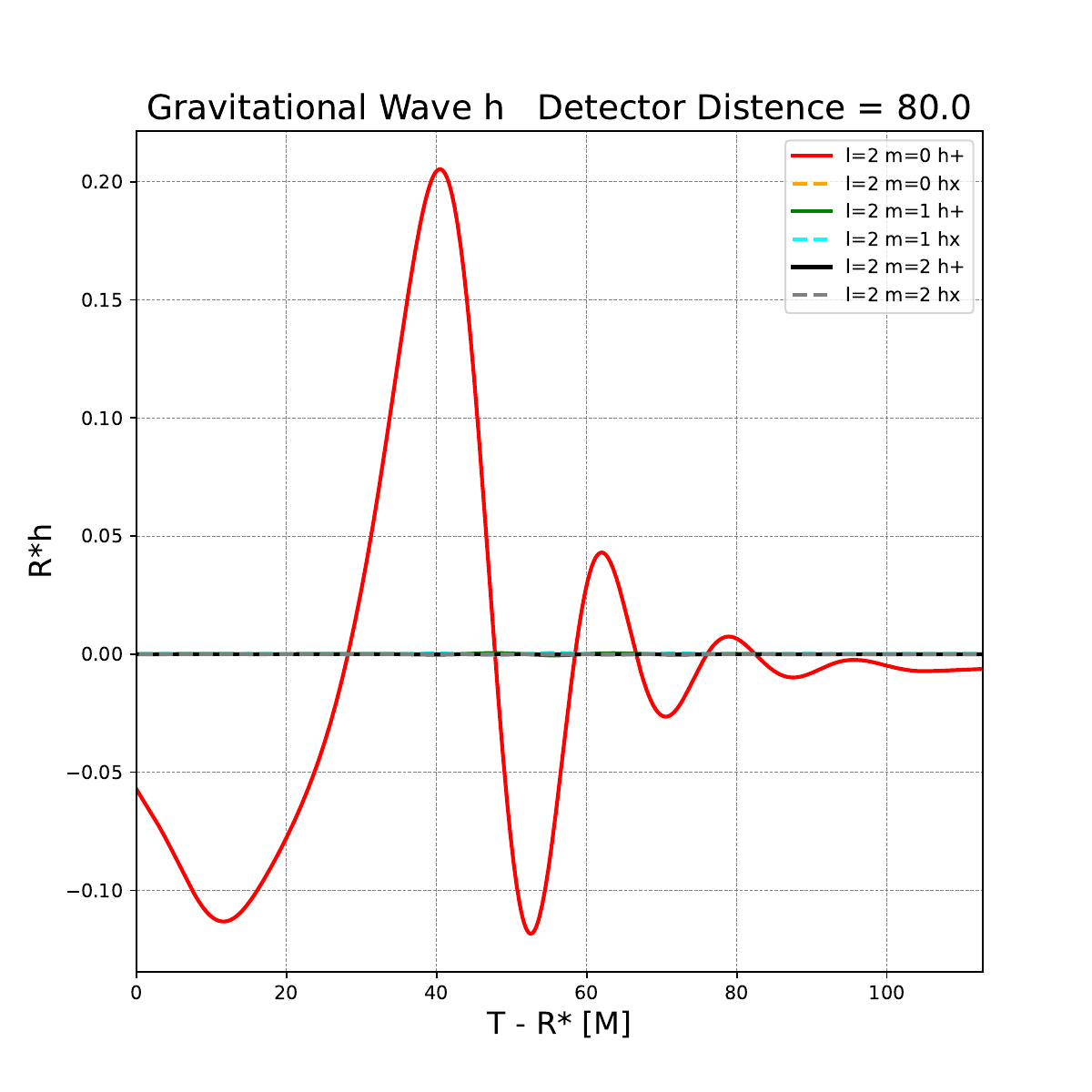}
	\caption{The gravitational waveform of the two polarization amplitudes $h_+$ and $h_\times$ for multiple black hole systems during the collision process. In this figure, the vertical axis labels the product $r \cdot h$, in which the waveform $h$ is extracted at the radius $r=80M$. The horizontal axis labels the retarded time $t_\ast = t - r_\ast$ calculated in terms of the Regge‑Wheeler tortoise coordinate (i.e., $r_\ast = r+2M\ln\left(\frac{r}{2M}-1\right)$).}
	\label{Fig BH GW Amplitude}
\end{figure*}

Figures \ref{Fig BH Trajectory}-\ref{Fig BH GW Amplitude} present the dynamical evolution of the black holes initially arranged in symmetric regular polygonal configurations, showing both the motions of the black holes and the gravitational waveforms emitted during the collision. For each case, the black holes' trajectories (confined to the $xy$ plane by symmetry), the gravitational waveform of the Weyl scalar $\Psi_{4}$ in the Newman-Penrose formalism, and the two gravitational-wave polarization amplitudes $h_{+}$, $h_{\times}$ extracted during the collision process are given in Figure \ref{Fig BH Trajectory}, Figure \ref{Fig BH Psi4}, and Figure \ref{Fig BH GW Amplitude}, respectively. In all these examples, the black holes collide at the center without experiencing any gravitational kick, and the emitted gravitational radiation is dominated by the $l=2$, $m=0$ mode of the plus polarization ($h_{+}$). The only exception is the case of the binary black hole collision, for which the $l=2$, $m=2$ mode also contributes substantially. This behavior indicates that the symmetry of black hole systems imposes stringent constraints on the modes of gravitational radiation. Furthermore, these numerical relativity results are consistent with the analytical predictions obtained under the conventional gravitational wave quadrupole approximation (see Appendix \ref{Appendix} for a detailed analysis).

\begin{figure*}
	\centering
	\includegraphics[width=0.325\textwidth]{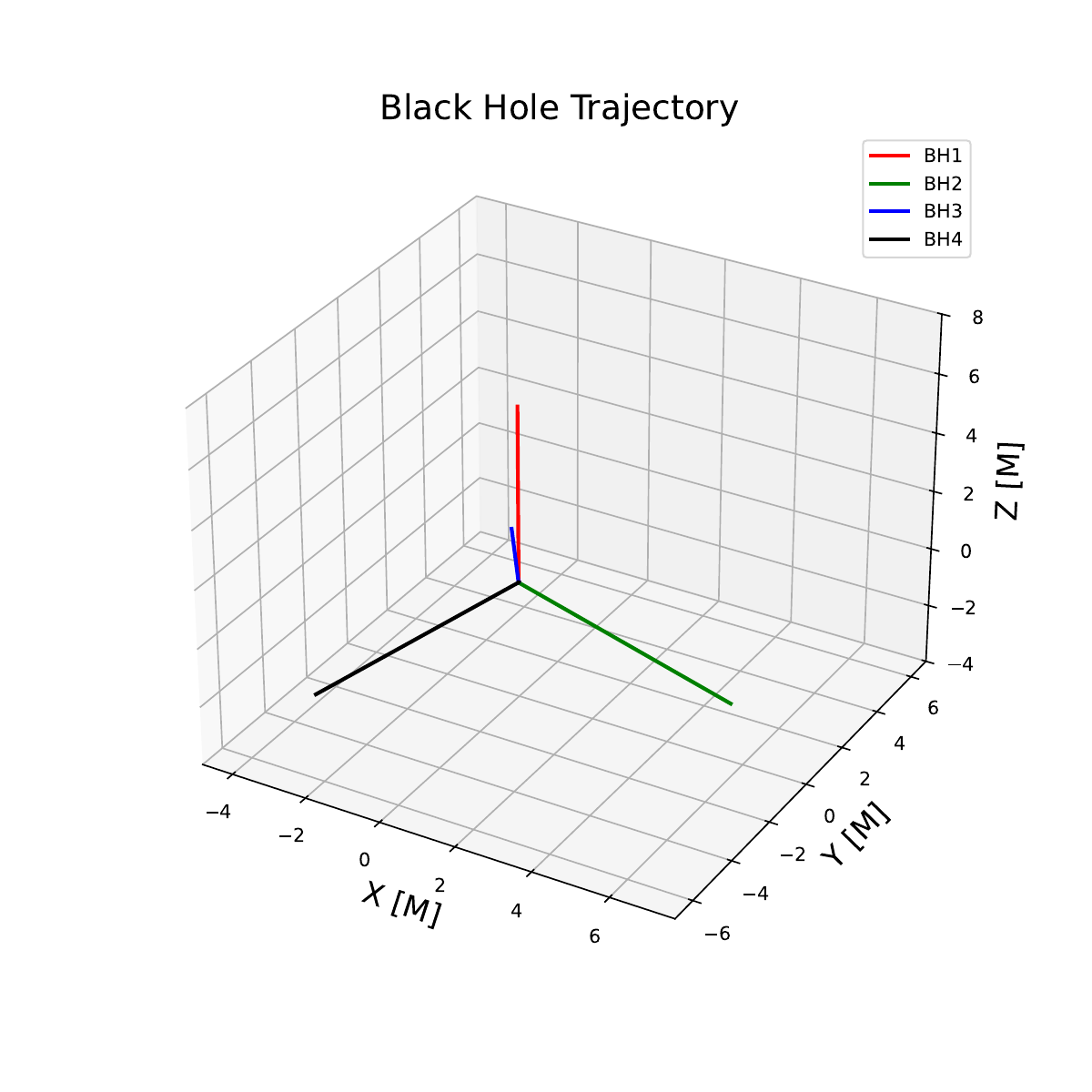}
	\includegraphics[width=0.325\textwidth]{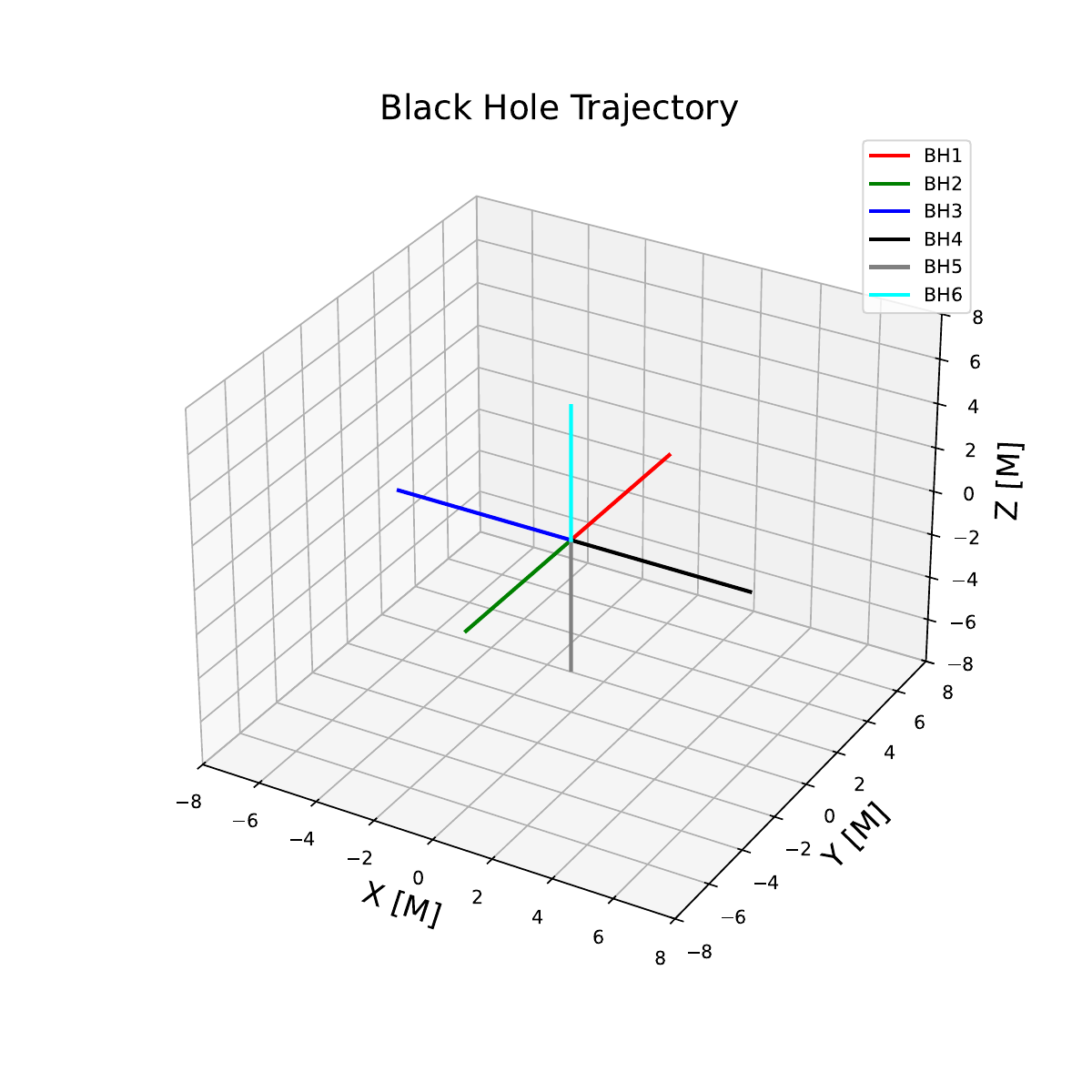}
	\includegraphics[width=0.325\textwidth]{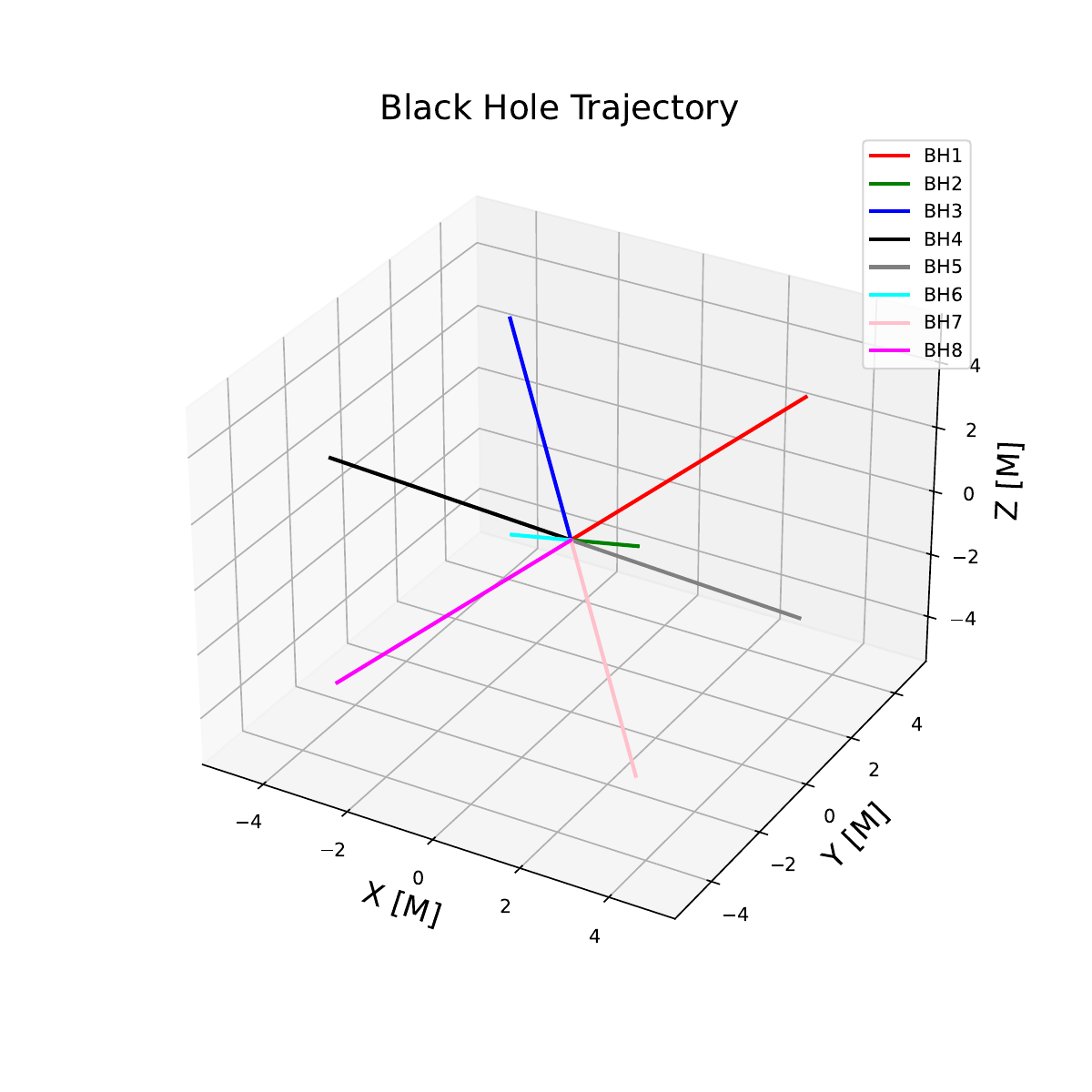}
	\includegraphics[width=0.325\textwidth]{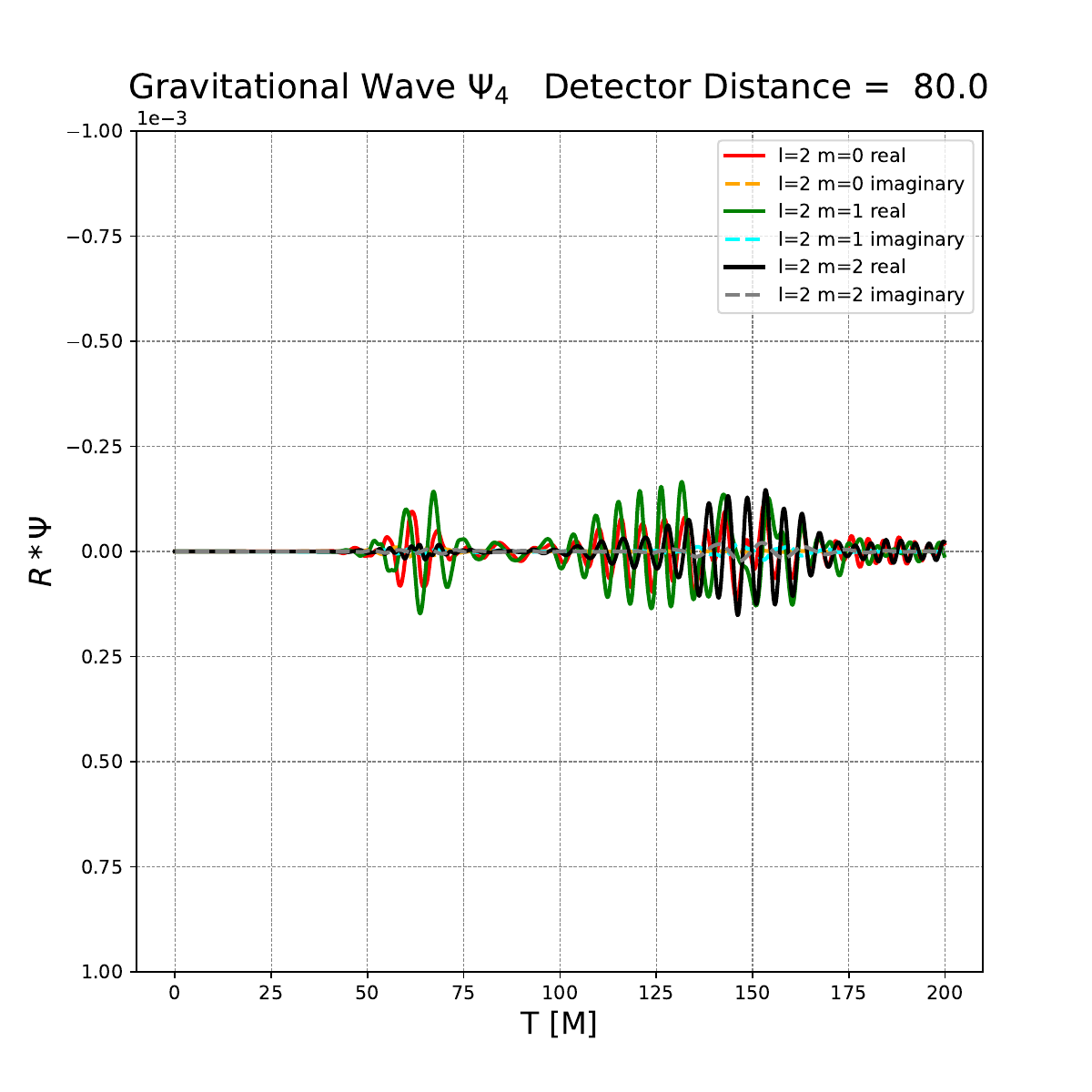}
	\includegraphics[width=0.325\textwidth]{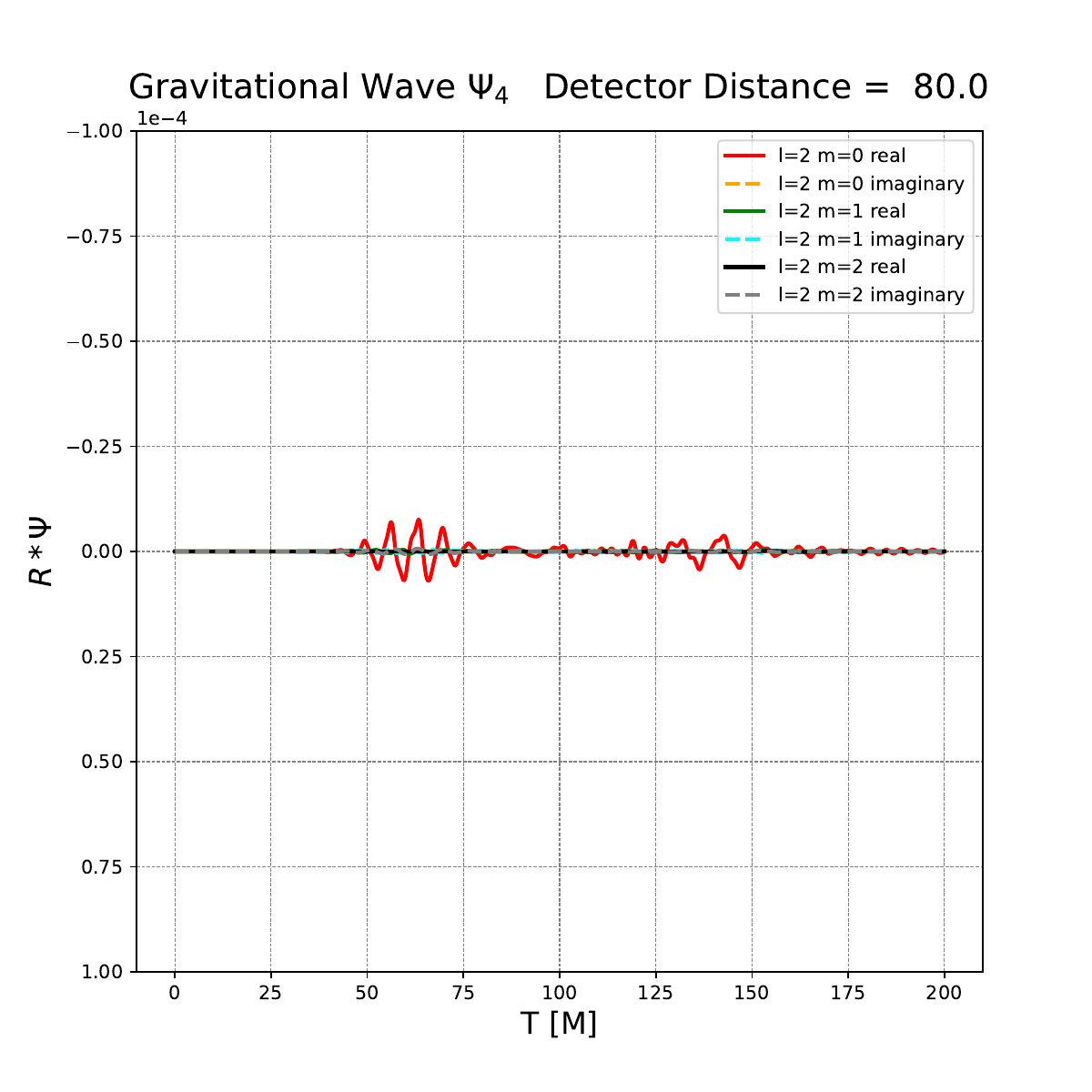}
	\includegraphics[width=0.325\textwidth]{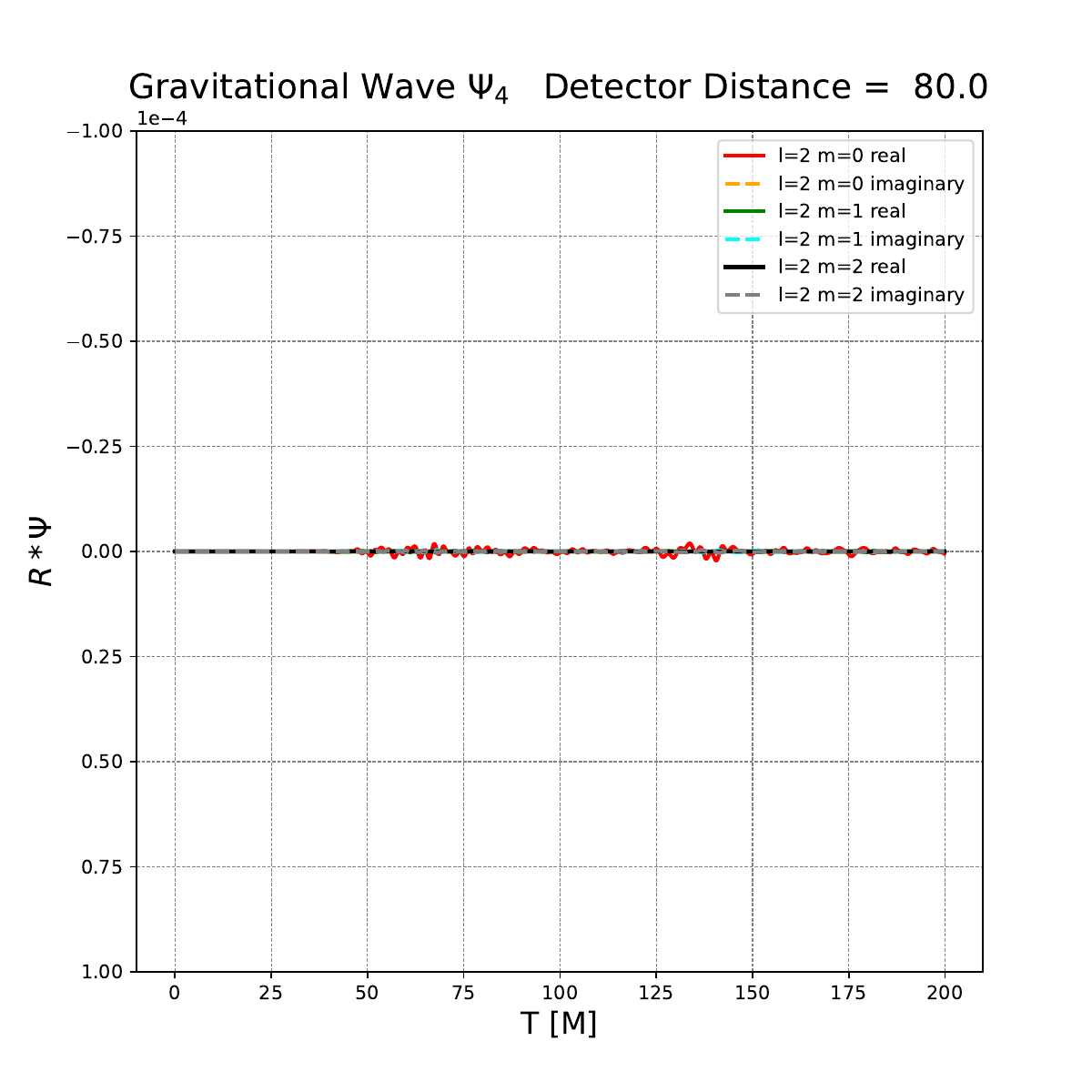}
	\includegraphics[width=0.325\textwidth]{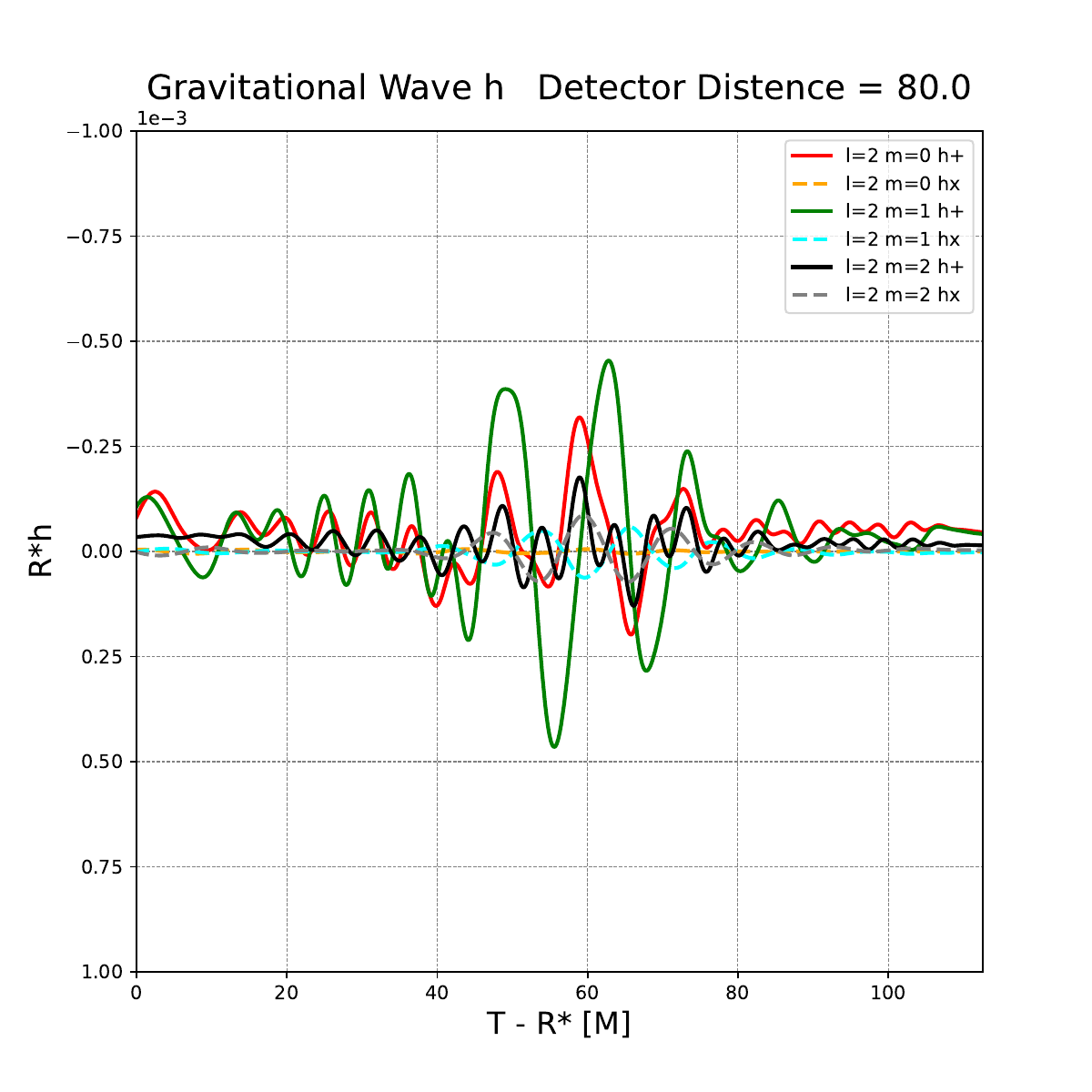}
	\includegraphics[width=0.325\textwidth]{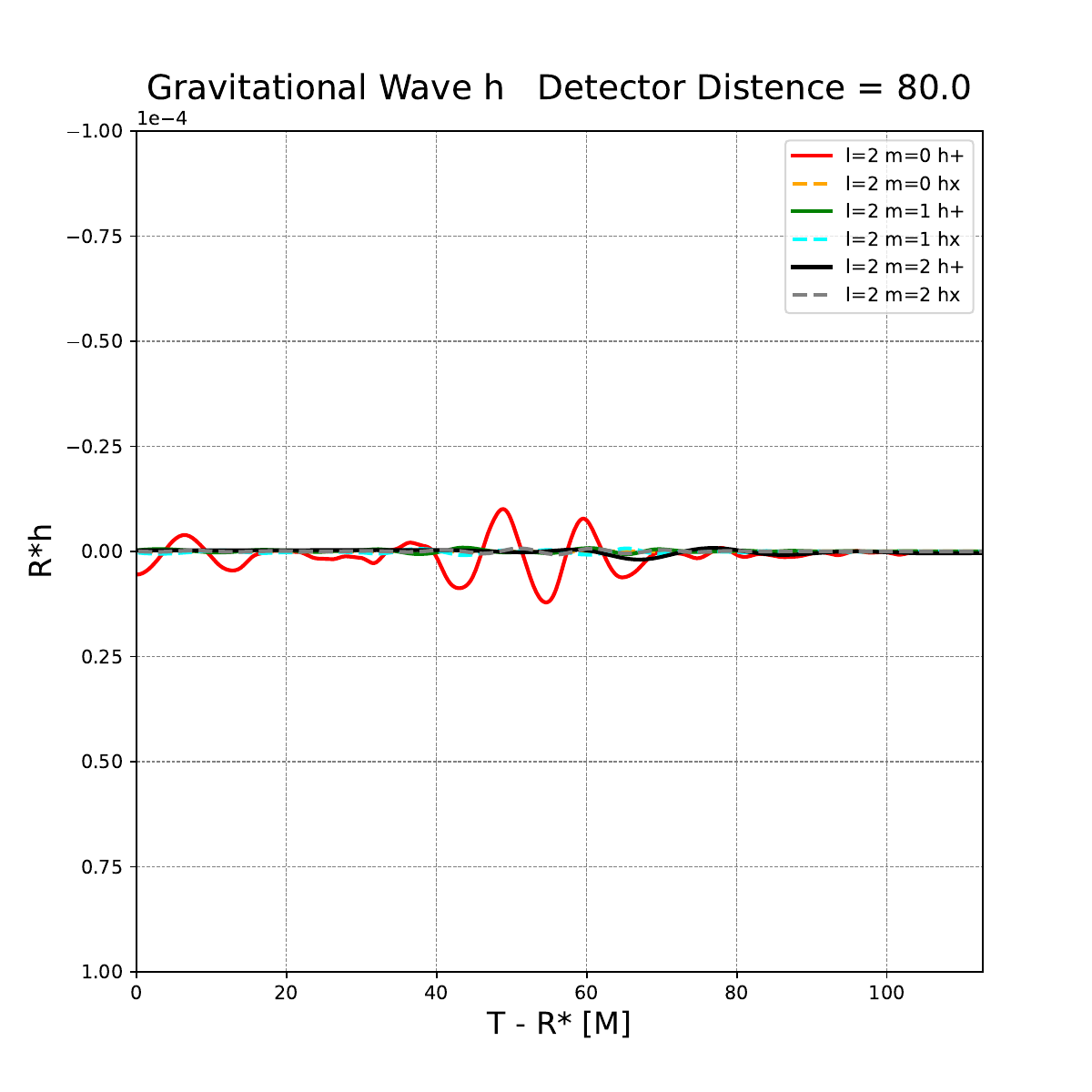}
	\includegraphics[width=0.325\textwidth]{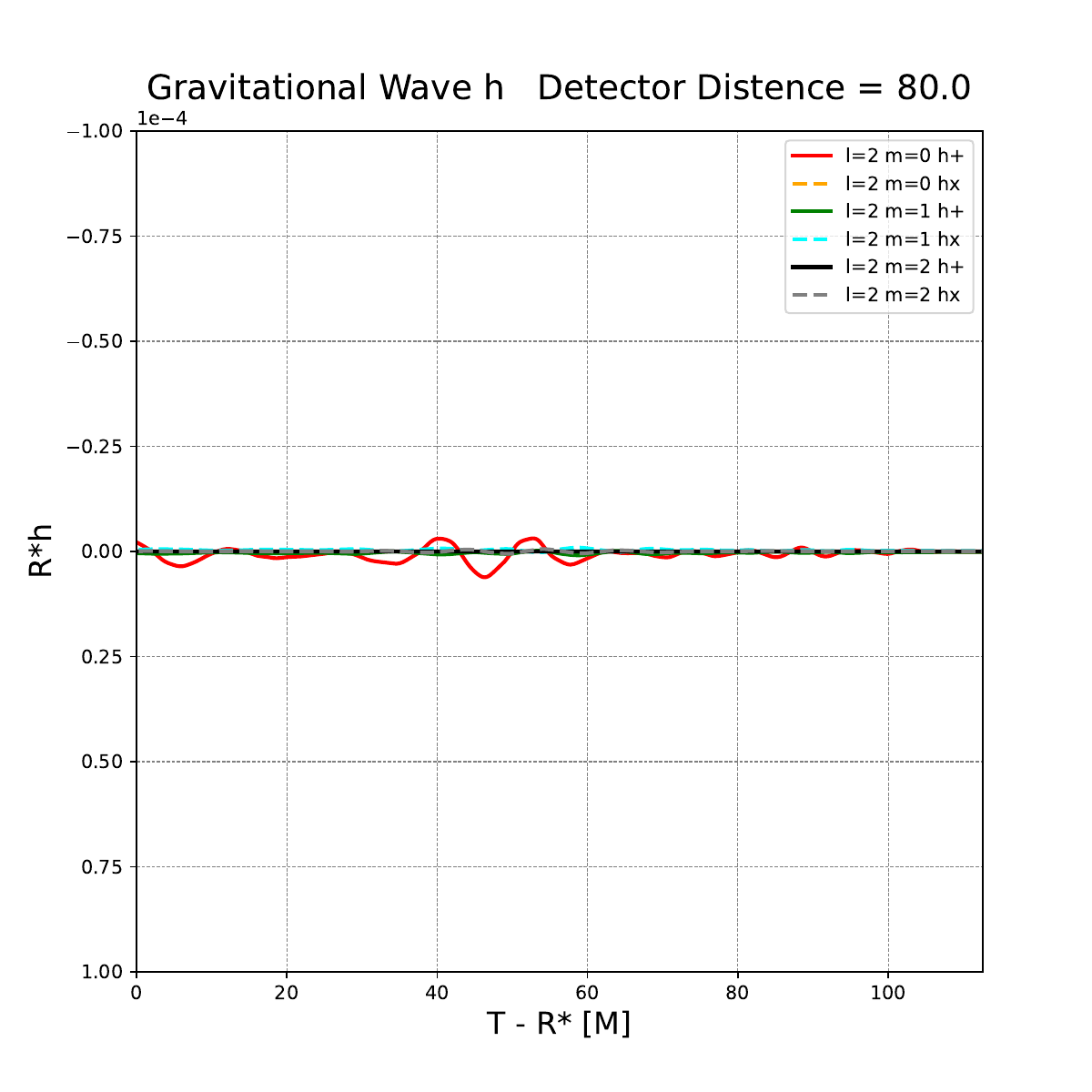}
	\caption{Head-on collisions of black holes in regular polyhedral configurations. From left to right: the regular tetrahedral configuration (4 black holes), the regular octahedral configuration (6 black holes), and the regular hexahedral (cubic) configuration (8 black holes). The upper, middle, lower rows shown the trajectories of black holes, the gravitational waveforms of the Weyl scalar $\Psi_4$, the waveform of the two polarization amplitudes $h_+$ and $h_\times$ during the collision process. The waveforms are extracted at the radius $r = 80M$.}
	\label{Fig Regular Polyhedron}
\end{figure*}

We next consider multi-black-hole collisions in another symmetric class of configurations: regular polyhedral configurations. Figure \ref{Fig Regular Polyhedron} illustrates the black hole collisions within several different regular polyhedral configurations: the regular tetrahedral configuration (4 black holes), regular octahedral configuration (6 black holes), and regular hexahedral configuration (i.e., the cubic configuration, 8 black holes).  
The black holes' 3-dimensional trajectories, the gravitational waveform of the Weyl scalar $\Psi_4$, and the gravitational wave polarization amplitudes $h_{+}$ and $h_{\times}$ are presented. In all these cases, the black holes collide at the center simultaneously, and the resulting merged black hole exhibits no gravitational recoil, which is in agreement with the regular polygonal cases shown in Figure \ref{Fig BH Trajectory}. Furthermore, it is worth noting that all multipole modes of the gravitational radiation are strongly suppressed in these regular polyhedral configurations (with the amplitudes of $\Psi_4$, $h_+$, and $h_\times$ being of order $10^{-4}$ or below). This suppression is significantly stronger than that observed in regular polygonal configurations (for regular polygonal configurations, only the $m \neq 0$ modes are suppressed). This enhanced suppression arises because the regular polyhedral configuration possesses higher symmetry than the regular polygonal configuration. These conclusions are consistent with the analytical predictions from the conventional gravitational wave quadrupole approximation, as presented in the Appendix \ref{Appendix}.

\begin{figure*}
	\centering
	\includegraphics[width=0.35\textwidth]{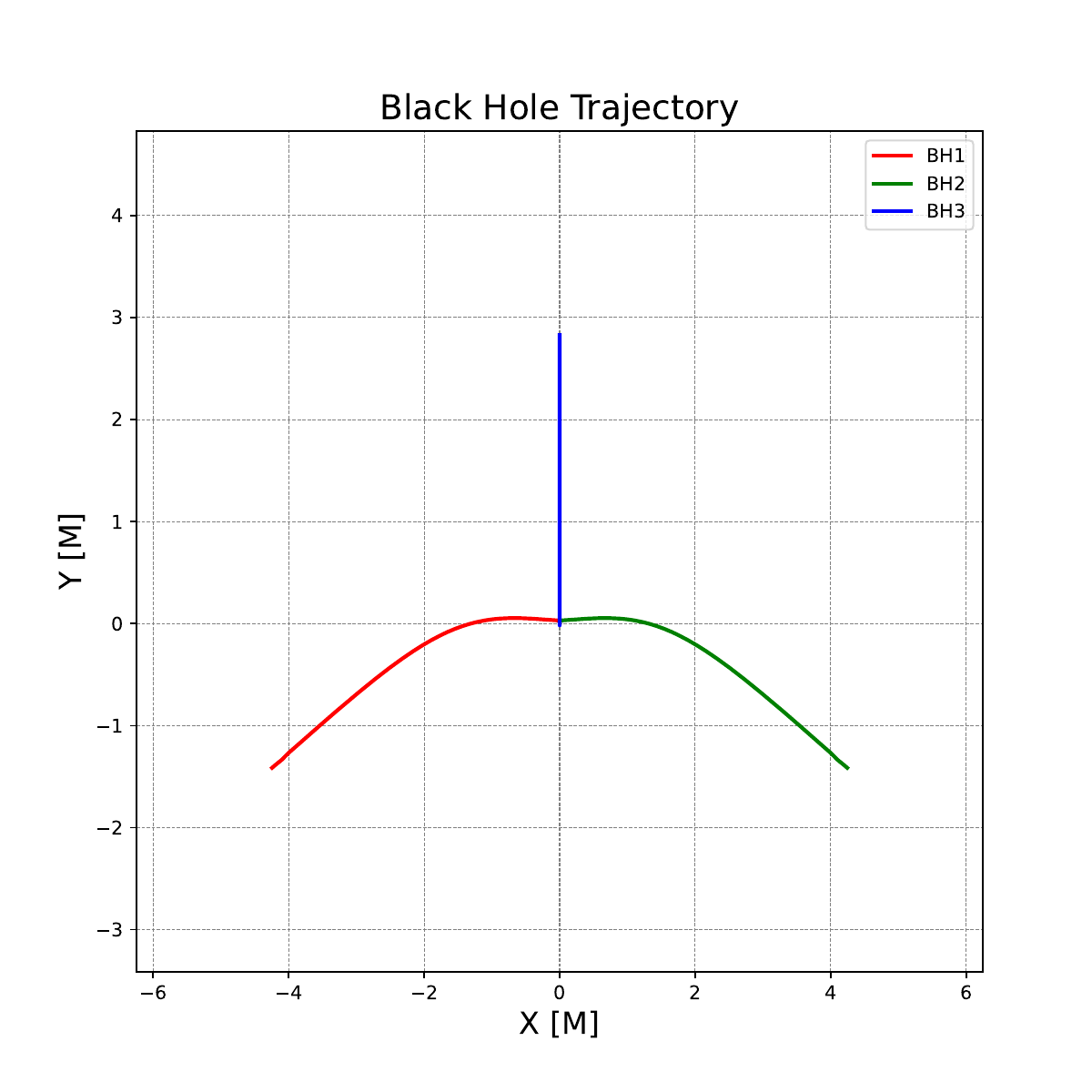}
	\includegraphics[width=0.35\textwidth]{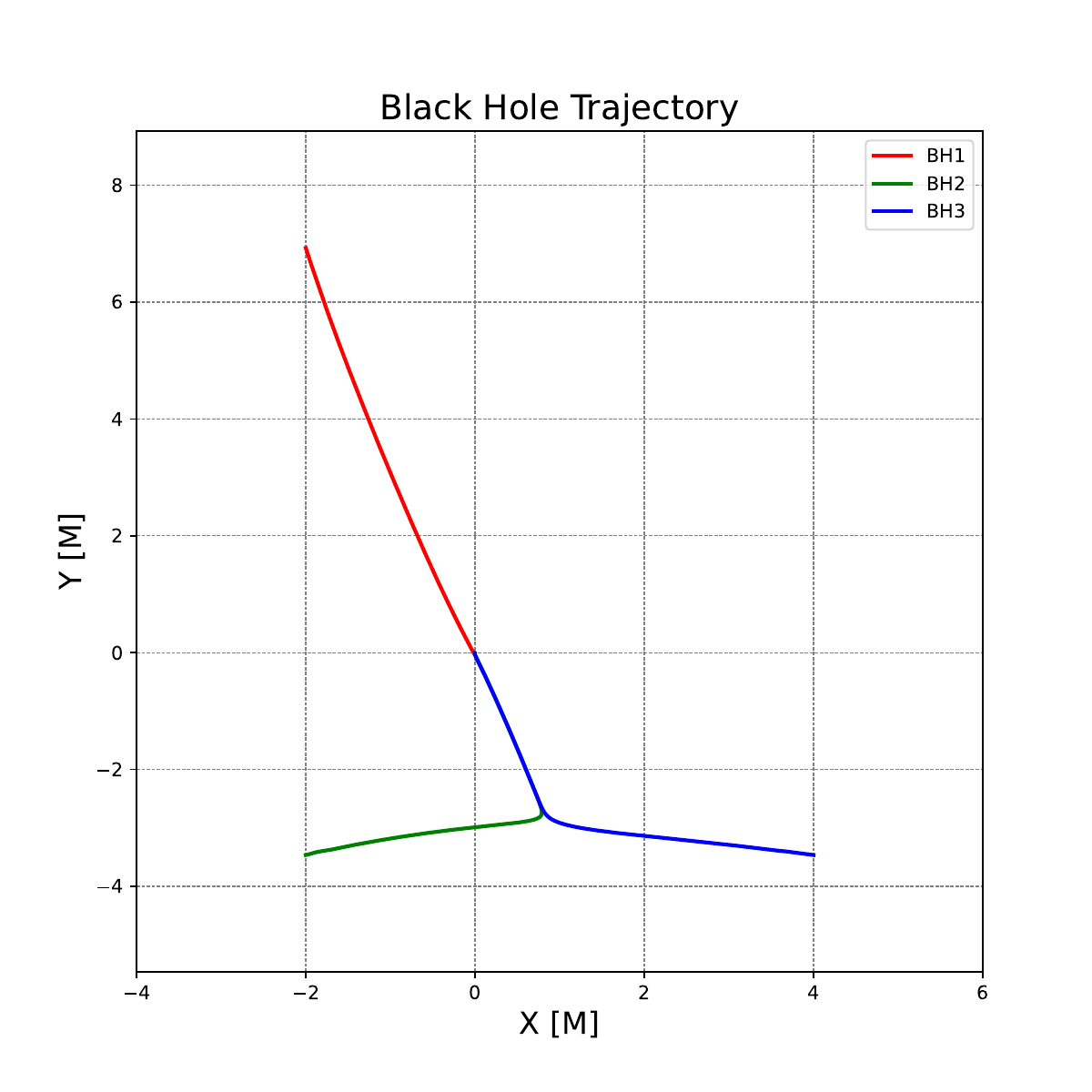}
	\includegraphics[width=0.35\textwidth]{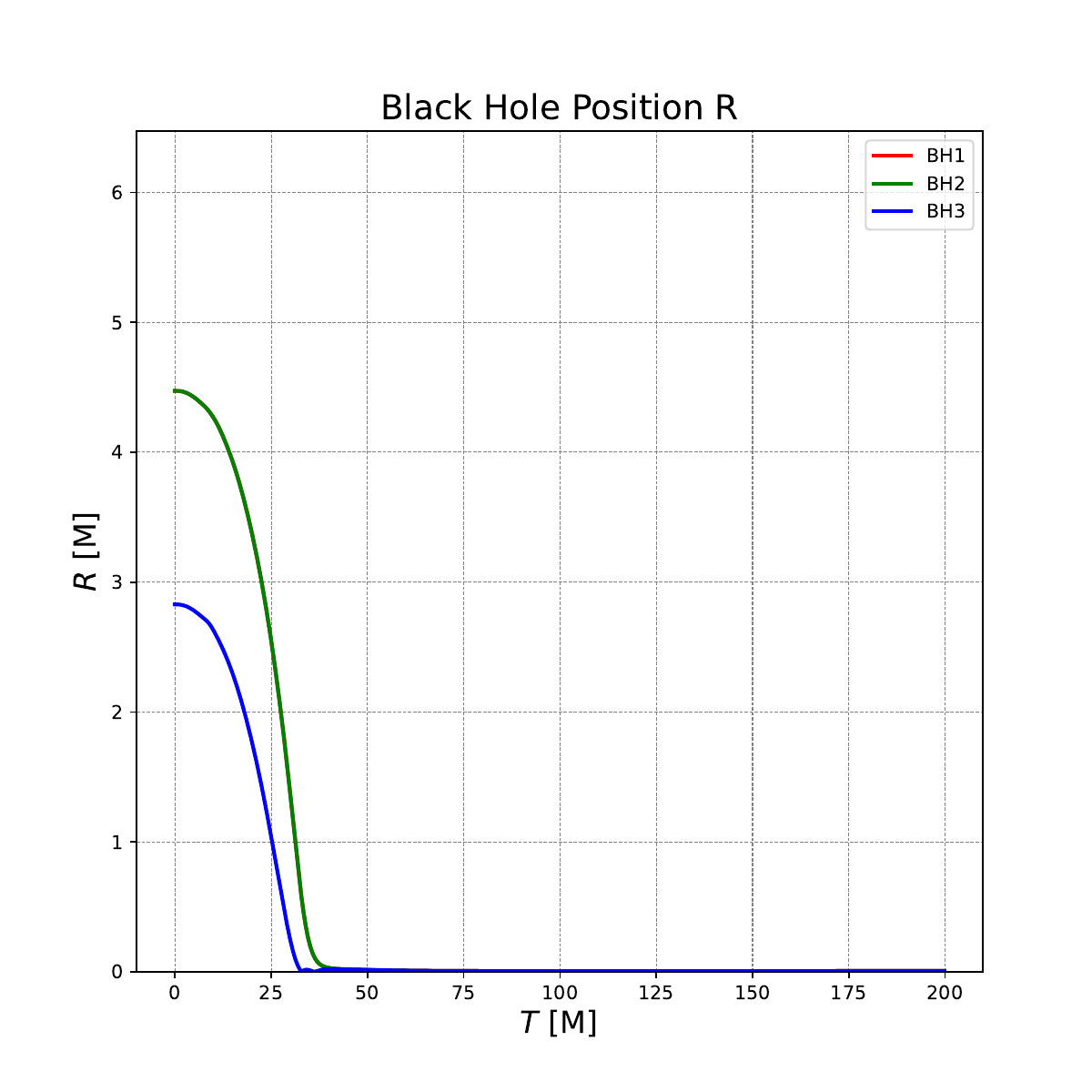}
	\includegraphics[width=0.35\textwidth]{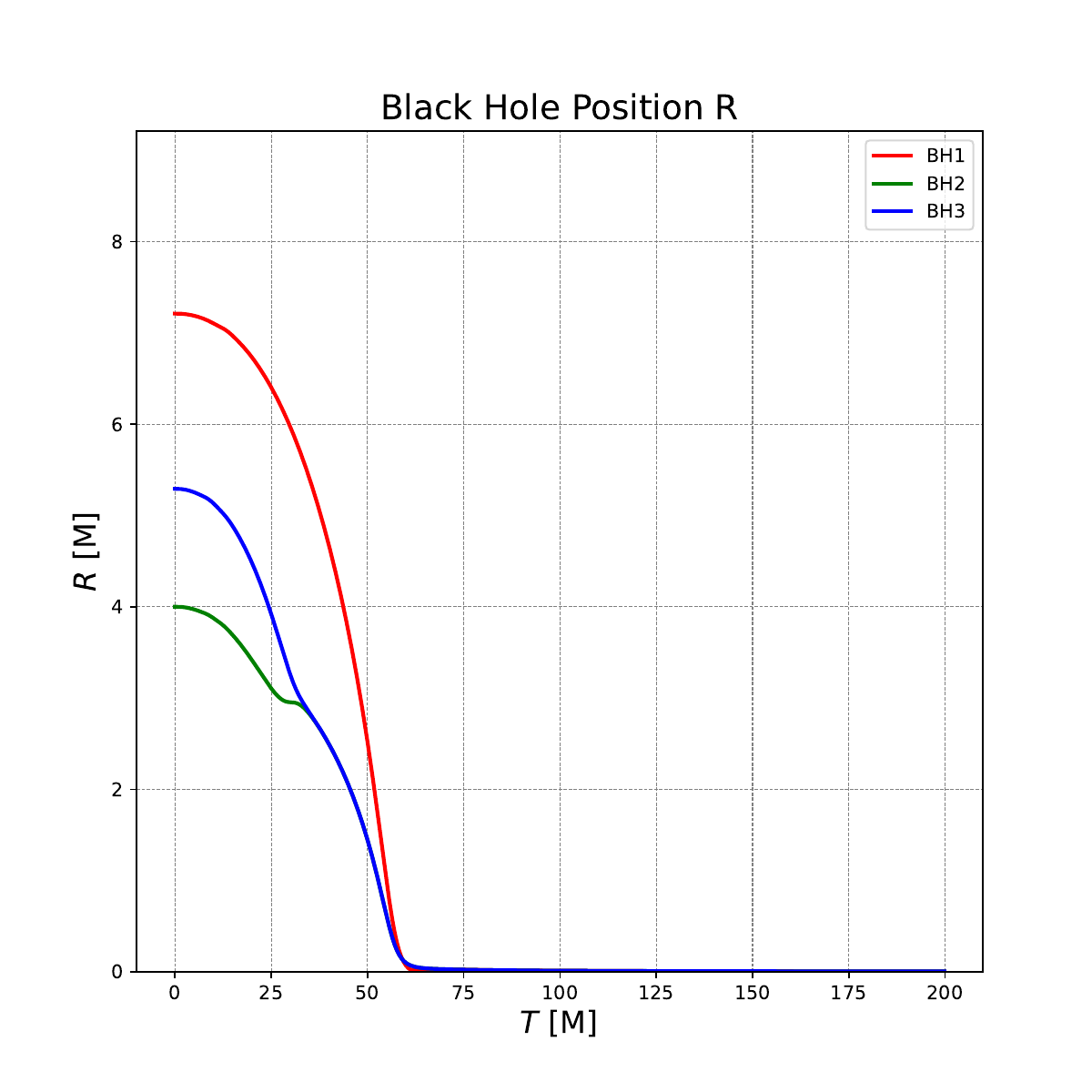}
	\includegraphics[width=0.35\textwidth]{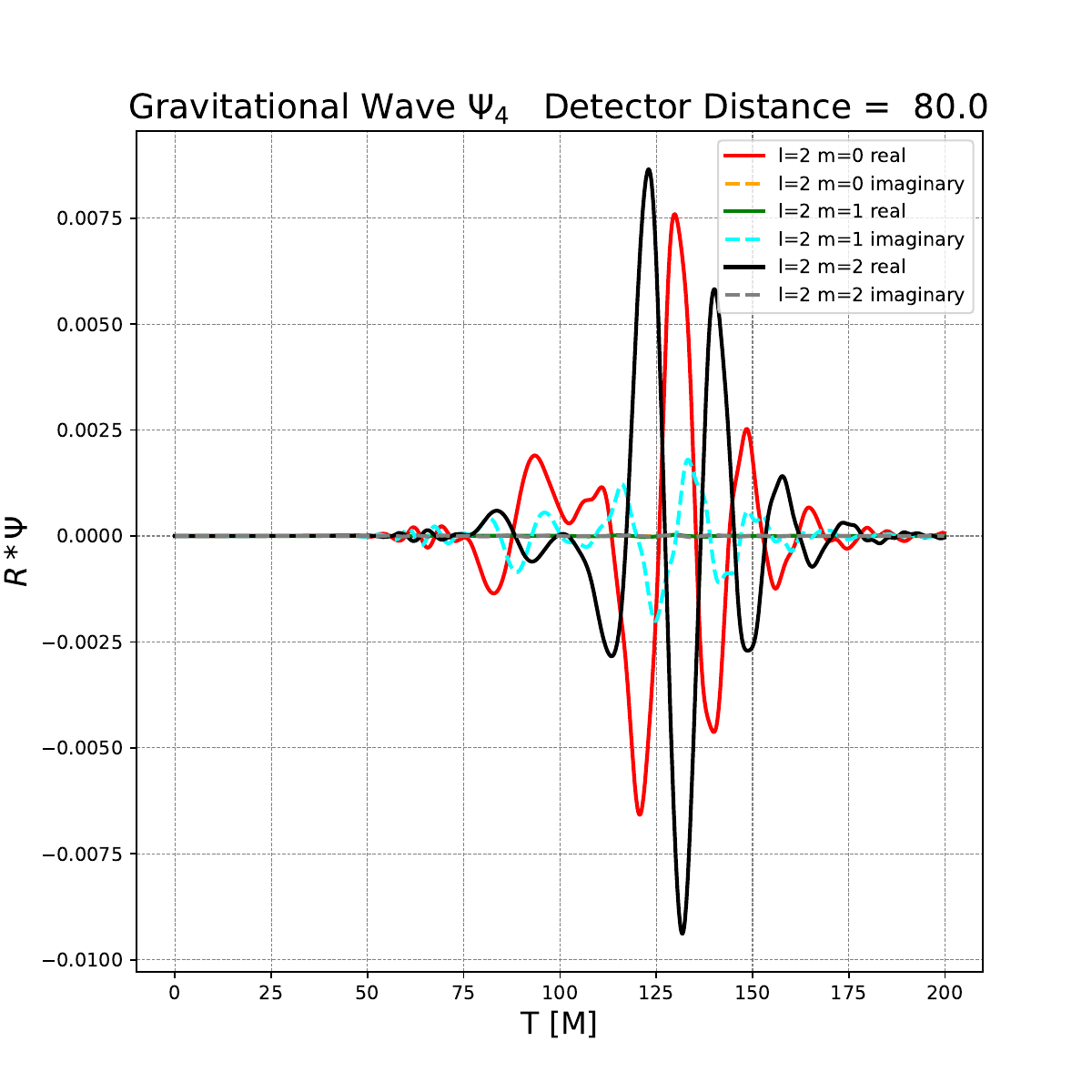}
	\includegraphics[width=0.35\textwidth]{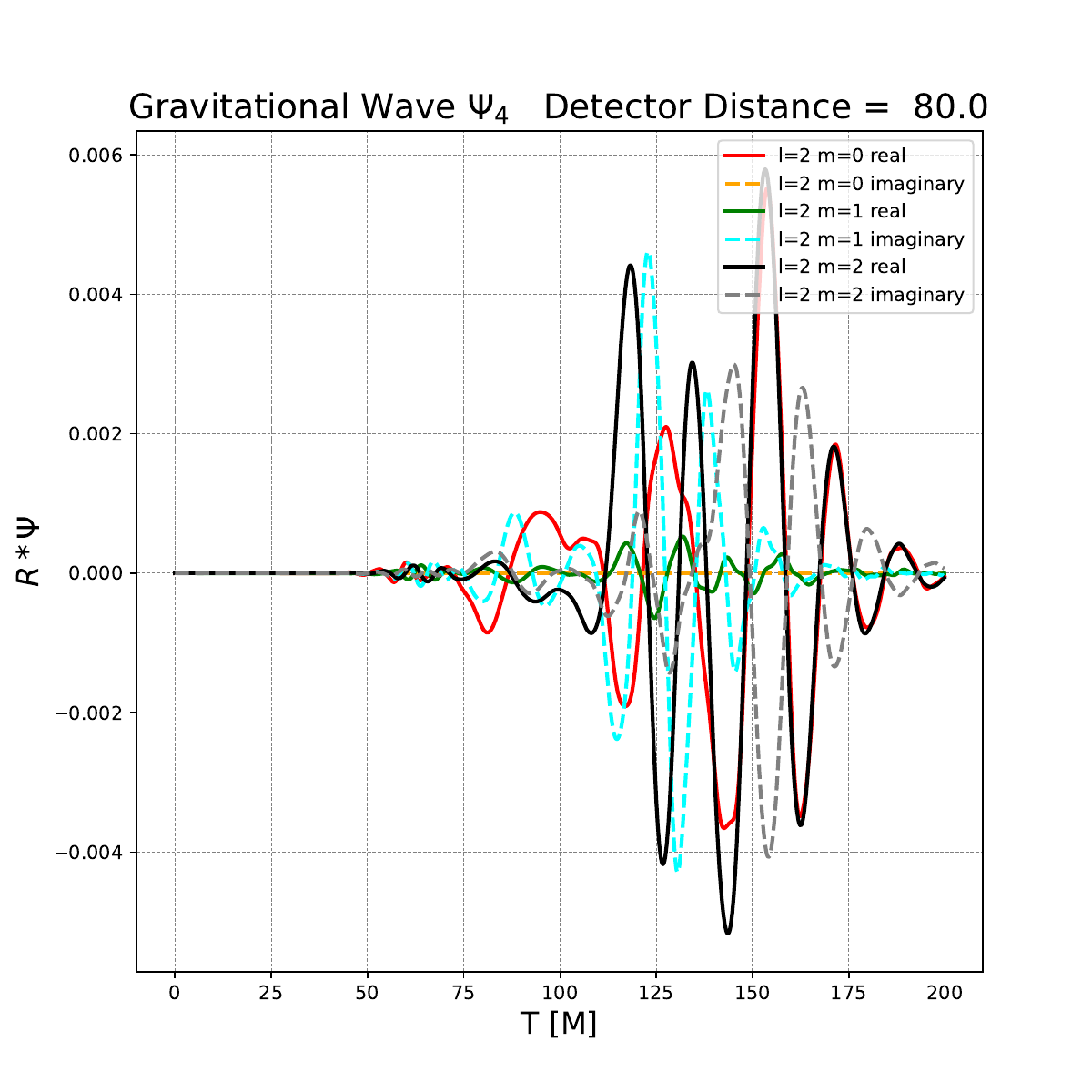}
	\includegraphics[width=0.35\textwidth]{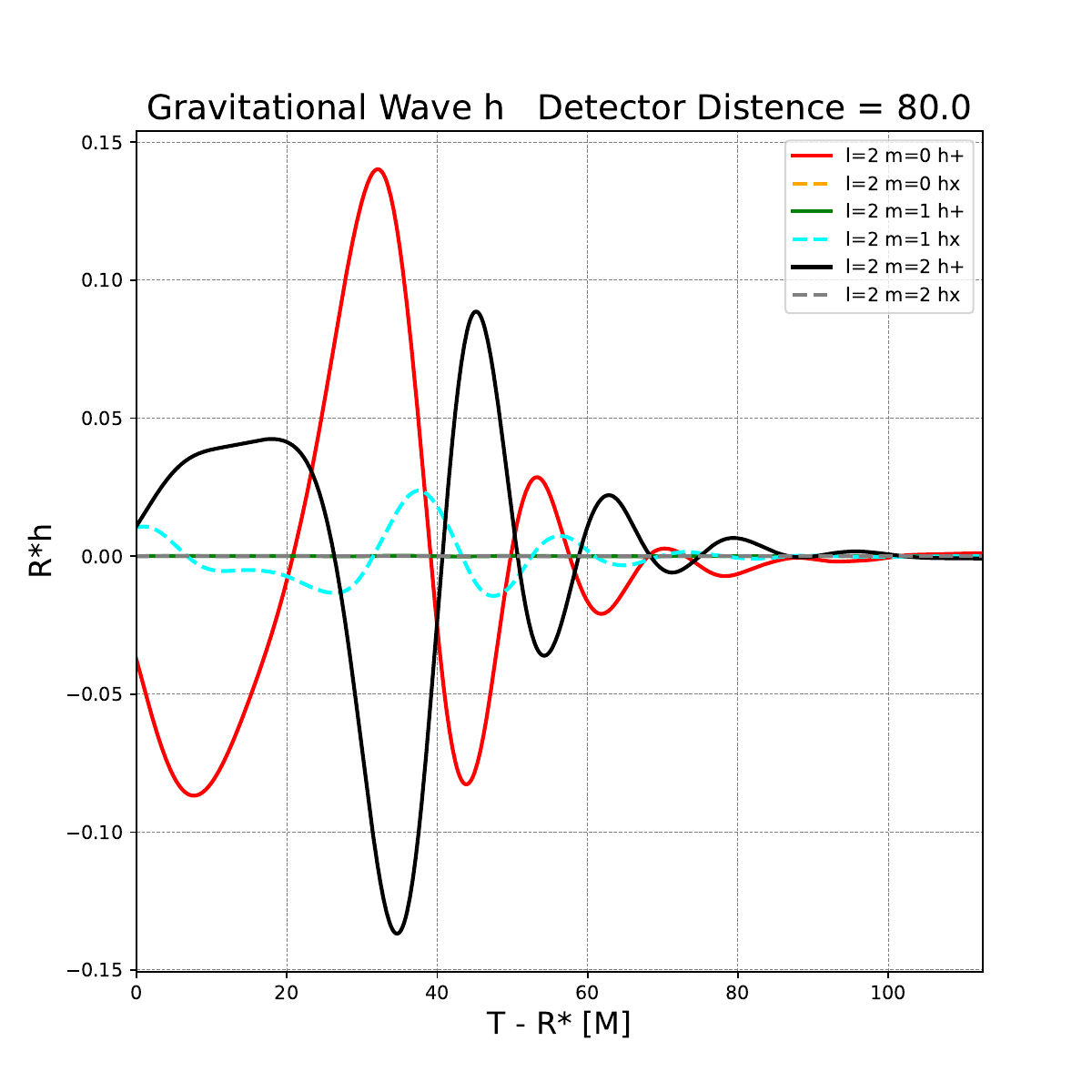}
	\includegraphics[width=0.35\textwidth]{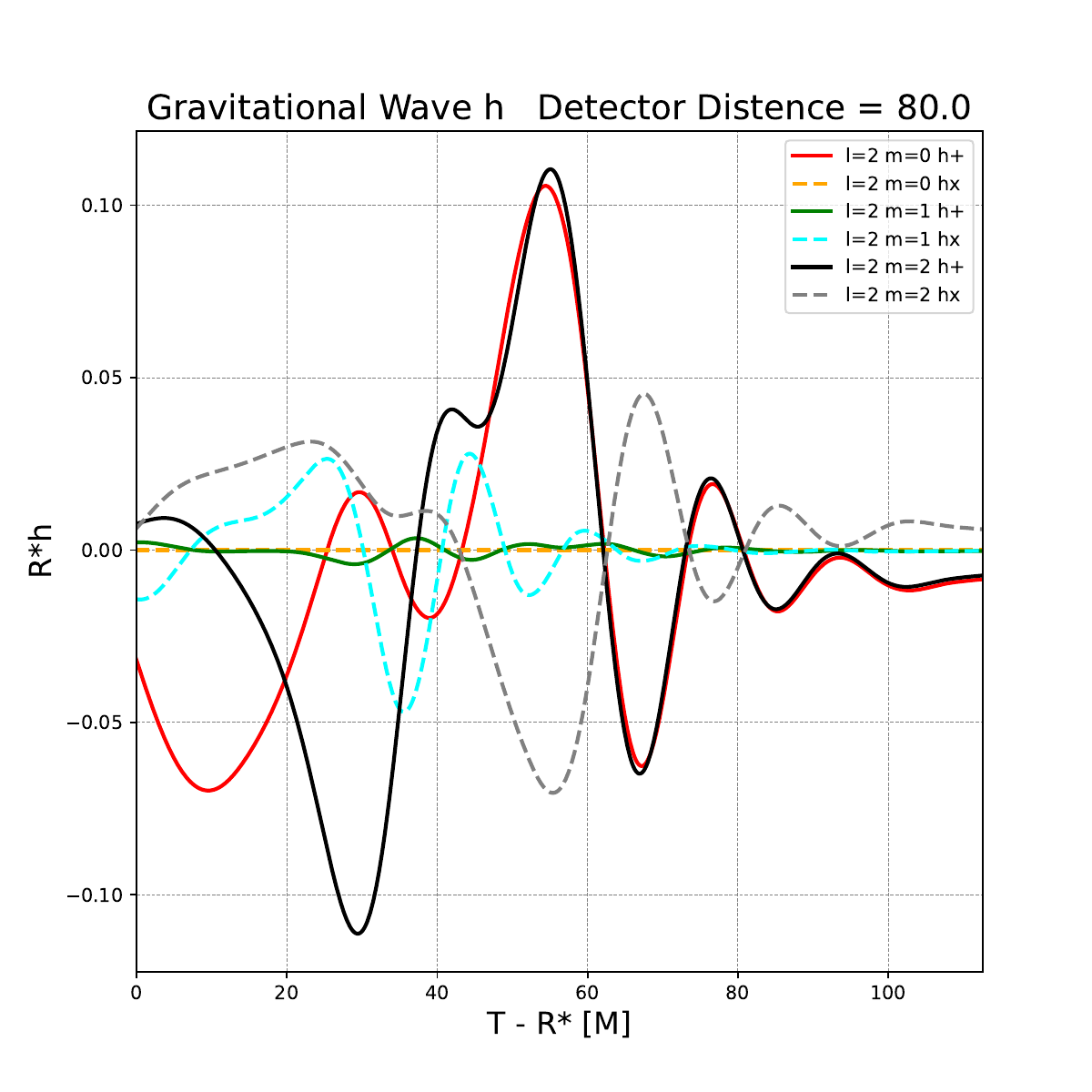}
	\caption{The dynamical evolution of three black hole systems in two non-symmetric triangular configurations. The left panel of this figure shows an isosceles right triangle configuration ($45^\circ$-$45^\circ$-$90^\circ$), and the right panel shows a right triangle configuration with angles of $30^\circ$ and  $60^\circ$. The first row gives the black holes' trajectories in the $xy$ plane. The second row gives the distance of each black hole to the center (i.e., $R_{I}=\sqrt{x_{I}^{2}+y_{I}^{2}+z_{I}^{2}}$). The last two rows present the gravitational waveforms of the Weyl scalar $\Psi_4$ and the gravitational waveform of the two polarization amplitudes $h_+$, $h_\times$ during the collision process (the waveforms are extracted at the radius $r = 80M$).}
	\label{Fig Non-symmetric Configurations}
\end{figure*}

Following the discussion of symmetric configurations, we now explore the gravitational wave characteristics of colliding black hole systems in non-symmetric configurations. Figure \ref{Fig Non-symmetric Configurations} displays the dynamical evolution of three-black-hole collisions and their gravitational waveforms in two non-symmetric triangular configurations. In each case, three black holes of equal mass ($m_{I}=\frac{M}{3}$) are initially positioned at the three vertices of a right triangle. Specifically, the left panel of Figure \ref{Fig Non-symmetric Configurations} shows an isosceles right triangle configuration ($45^\circ$-$45^\circ$-$90^\circ$), while the right panel depicts a right triangle configuration with angles of $30^\circ$ and  $60^\circ$. In both cases, the remnant black hole remains at the center after the collision, experiencing no gravitational kick. However, the three black holes do not necessarily collide simultaneously (see the second row of Figure \ref{Fig Non-symmetric Configurations}). In the left panel, the black hole located above the $x$-axis (shown in the blue curve) first moves to the center before colliding with the other two black holes. In the right panel, the two black holes located below the $x$-axis (corresponding to blue and green curves) first collide with each other at position $(x,y) \approx (0.8M, -2.8M)$, then subsequently move toward the center together with the third black hole, eventually merging at the center. Moreover, we observe that in these asymmetric configurations, all gravitational wave modes contribute significantly, rather than being restricted to the $l=2$, $m=0$ mode. This indicates that the asymmetry of the black hole system substantially enhances the gravitational radiation with $m \neq 0$.

Comparing the numerical results for black hole collisions in symmetric and non-symmetric configurations reveals that the symmetry of the multiple black hole system imposes strong constraints on gravitational radiation. A higher degree of spatial symmetry suppresses the multipolar richness of the emitted gravitational waves, particularly the contributions of the $m \neq 0$ modes, whereas breaking the symmetry substantially enhances the gravitational wave emission in these modes.

\section{Conclusion and Prospects}\label{sec4}

In this work, we have carried out a systematic study of the head-on collisions of multiple black hole systems in the framework of numerical relativity. The dynamical evolution of black hole systems is obtained by solving the Cauchy initial value problem constructed from the ADM $3+1$ decomposition of Einstein's field equations, and the numerical implementation is performed through the BSSN formulation with Bowen-York type initial data. Our simulations are carried out using the \texttt{AMSS-NCKU} code. By comparing black hole systems with symmetric and non-symmetric initial configurations, we have examined in detail how the symmetry of the black hole system governs the multipole modes of the resulting gravitational radiation.

For the symmetric configurations -- regular polygonal and regular polyhedral configurations -- the black holes always collide simultaneously at the center, and the merged black hole exhibits no gravitational recoil, as expected from the symmetry of the system. Furthermore, the gravitational radiation is strongly suppressed in these cases. For polygonal configurations, the emission is dominated by the $l=2$, $m=0$ mode, while modes with $m \neq 0$ are suppressed. For polyhedral configurations, the suppression is even stronger: all multipole modes are damped to amplitudes of order $10^{-4}$ or lower, owing to the higher degree of symmetry of the system. These results are in excellent agreement with the analytical predictions from the conventional gravitational wave quadrupole approximation.

In contrast, for the non-symmetric configurations, such as the right triangular configurations presented in Figure \ref{Fig Non-symmetric Configurations}, the collision dynamics become more intricate: the black holes no longer merge simultaneously, and intermediate collisions can occur away from the center before the final merger. Correspondingly, because of the asymmetry of the system, the gravitational radiation is significantly enriched, with the $m \neq 0$ modes making substantial contributions to the total gravitational wave emission.

In conclusion, our results demonstrate that the symmetry of multiple black hole systems can impose stringent constraints on the allowed gravitational wave modes, while breaking this symmetry not only modifies the collision sequence but also provides an effective mechanism for enhancing gravitational wave emission, particularly for the modes with $m \neq 0$. These findings may offer useful guidance for understanding the gravitational wave signatures of more general relativistic $N$-body systems and black hole encounters. In future work, it would be interesting to extend this study to black hole systems with unequal masses or nonzero spins, in order to explore the interplay between symmetry breaking and gravitational wave signatures in more realistic astrophysical $N$-body scenarios.

\appendix
\section{Analysis From the Quadrupole Approximation} \label{Appendix}

In this appendix, we provide an analysis of the gravitational wave amplitudes for black hole collisions within the quadrupole approximation. Such an analysis provides an interpretation to better explain our numerical relativity results presented in Section \ref{sec3}.

In the quadrupole approximation, the spatial components of the metric perturbation in the transverse-traceless gauge are given by the second time derivative of the quadrupole moment tensor 
\begin{equation}
	h_{ij}^{\text{TT}} \sim \frac{2}{r} \cdot \frac{d^{2}\mathcal{I}_{ij}}{dt^{2}}\bigg|_{t_{\text{retard}}=t-r} 
	= \frac{2}{r} \cdot \ddot{\mathcal{I}}_{ij}(t-r) .
\end{equation}
The conventional quadrupole moment tensor is defined by
\begin{equation}
	\mathcal{I}_{ij}(t) = \int \rho(x,t) \bigg( x_{i}x_{j} - \frac{1}{3}  \delta_{ij} r^{2} \bigg)  d^{3} x .
\end{equation}
In the far-field regime when $r$ becomes sufficiently large, the gravitational wave admits the following multipole expansion
\begin{eqnarray}
	h(t,r,\theta,\phi) & = & \sum_{l=2}^{\infty} \sum_{m=-l}^{l} h_{lm}(t,r) \cdot  Y_{s=-2}^{lm}(\theta,\phi) \nonumber 
	\\
	& = & \sum_{l=2}^{\infty} \sum_{m=-l}^{l} \bigg[ h^{+}_{lm}(t,r) + i h^{\times}_{lm}(t,r) \bigg] \cdot Y_{s=-2}^{lm}(\theta,\phi) .
\end{eqnarray}
In the multipole expansion, if we choose the polar axis to coincide with the coordinate $z$-axis, then the multipole expansion of gravitational wave in the quadrupole approximation gives \cite{Multipole-Thorne-1980}
\begin{subequations}
	\begin{eqnarray}
		h_{20} = h_{20}^{+} + i h_{20}^{\times}
		\ &\sim& \ \ddot{\mathcal{I}}_{zz} \label{h20} ,
		\\
		h_{21} = h_{21}^{+} + i h_{21}^{\times}
		\ &\sim& \ - (\ddot{\mathcal{I}}_{xz} - i \ddot{\mathcal{I}}_{yz}) \label{h21} ,
		\\
		h_{22} = h_{22}^{+} + i h_{22}^{\times}
		\ &\sim& \ (\ddot{\mathcal{I}}_{xx} - \ddot{\mathcal{I}}_{yy}) - 2 i \ddot{\mathcal{I}}_{xy} \label{h22} .
	\end{eqnarray}
\end{subequations}

When we consider a gravitational system with a regular polygonal or regular polyhedral configuration, each celestial body can be approximated as a point mass. In such cases, the matter distribution of each celestial body reduces to $\rho_{I}(\mathbf{r})=m_{I}\delta(\mathbf{r}-\mathbf{r_{I}})$, with $m_{k}$ and $\mathbf{r_{k}}$ being the mass and position of the celestial body. In this way, the quadrupole moment components for the gravitational system can be calculated as
\begin{equation}
	\mathcal{I}_{ij}(t) = \sum_{I=1}^{N} m_{I} \bigg[ x_{i}^{I} x_{j}^{I} - \frac{1}{3}  \delta_{ij} r_{I}^{2}(t) \bigg] ,
	\label{quadrupole moment component}
\end{equation}
where $x_{i}^{I}$ is the $i$-th coordinate of the $I$-th celestial body, $m_{I}$ is the mass of this celestial body, and $r_{I}$ is its distance to the center.

\subsection{Regular Polygonal Configurations} \label{Appendix Polygon}

Particularly, for the regular polygonal configuration, each black hole is initially located at a vertex of the regular polygon in the $xy$ plane and moves radially towards the center. Its position vector is
\begin{equation}
	\mathbf{r}(t) = (R(t)\cos{\phi_{I}}, R(t)\sin{\phi_{I}}, 0) ,
\end{equation}
where $R(t)$ denotes the common distance of the black holes from the center (it is identical for all black holes, as required by the symmetry of the configuration). From this position vector, the quadrupole moment components in equation (\ref{quadrupole moment component}) can be calculated
\begin{subequations}
\begin{eqnarray}
	\mathcal{I}_{xx}(t) & = & \sum_{I=1}^{N} m_{I} \bigg[ x_{I}^{2}(t) -\frac{1}{3}R^{2}(t) \bigg] 
	= \sum_{I=1}^{N} m_{I} R^{2}(t) \bigg( \cos^{2}\phi_{I} - \frac{1}{3} \bigg) , \label{quadrupole moment Ixx} \\
	\mathcal{I}_{yy}(t) & = & \sum_{I=1}^{N} m_{I} \bigg[ y_{I}^{2}(t) - \frac{1}{3} R^{2}(t) \bigg]
	= \sum_{I=1}^{N} m_{I} R^{2}(t) \bigg( \sin^{2}\phi_{I} -\frac{1}{3} \bigg) , \label{quadrupole moment Iyy} \\
	\mathcal{I}_{zz}(t) & = & \sum_{I=1}^{N} m_{I} \bigg[ z_{I}^{2}(t) - \frac{1}{3} R^{2}(t) \bigg] 
	= - \sum_{I=1}^{N} \frac{1}{3} m_{I} R^{2}(t) , \label{quadrupole moment Izz}  \\
	\mathcal{I}_{xy}(t) & = & \sum_{I=1}^{N} m_{I} \cdot x_{I}(t) \cdot y_{I}(t) = \sum_{I=1}^{N} m_{I} R^{2}(t) \sin\phi_{I} \cos\phi_{I} , \label{quadrupole moment Ixy} \\
	\mathcal{I}_{xz}(t) & = & \sum_{I=1}^{N} m_{I} \cdot x_{I}(t) \cdot z_{I}(t) = 0 , \\
	\mathcal{I}_{yz}(t) & = & \sum_{I=1}^{N} m_{I} \cdot y_{I}(t) \cdot z_{I}(t) = 0 .
\end{eqnarray}
\end{subequations}
Meanwhile, since the motion of each black hole is purely radial, its azimuthal angle remains unchanged during the entire collision process and is given by
\begin{equation}
	\phi_{I} = \phi_{0} + \frac{2\pi(I-1)}{N},
	\qquad I = 1, 2, \dots, N .
	\label{eq:vertices}
\end{equation}
This regular polygon arrangement implies a useful identity: for this $N$ black hole system, the following complex geometric series sums to zero
\begin{equation}
	\sum_{I=1}^{N} e^{2i\phi_I}
	= e^{2i\phi_0}\sum_{I'=0}^{N-1} e^{4\pi i \cdot \frac{I'}{N}}
	= \frac{e^{2i\phi_0}\,\bigl(e^{4\pi i}-1\bigr)}{e^{\frac{4\pi i}{N}}-1}
	= 0,
	\label{eq:angular_sum}
\end{equation}
where the second equality follows from the geometric-series formula. The numerator vanishes since $e^{4\pi i}=1$, while the denominator is kept nonzero unless $e^{\frac{4\pi i}{N}} = 1$ (which would require $N = 2$). Hence the sum vanishes identically for all $N\ge3$. Taking the real and imaginary parts of
equation (\ref{eq:angular_sum}) and using the half-angle identities, we immediately obtain
\begin{equation}
	\sum_{I=1}^{N} \cos^2\phi_{I} = \sum_{I=1}^{N} \sin^2\phi_{I} = \frac{N}{2},
	\qquad\quad
	\sum_{I=1}^{N} \sin\phi_{I} \cdot \cos\phi_{I}
	= \frac{1}{2}\sum_{I=1}^{N} \sin 2\phi_I = 0.
	\label{eq:trig_sums}
\end{equation}
Substituting equation (\ref{eq:trig_sums}) into equations (\ref{quadrupole moment Izz})-(\ref{quadrupole moment Ixy}), we obtain
\begin{subequations}
\begin{eqnarray}
	&& \mathcal{I}_{xx} = \mathcal{I}_{yy} = \frac{M R^{2}(t)}{6} , \\
	&& \mathcal{I}_{zz} = - \frac{M R^{2}(t)}{3} , \\
	&& \mathcal{I}_{xy} = \mathcal{I}_{xz} = \mathcal{I}_{yz} = 0 ,
\end{eqnarray}
\end{subequations}
where $M \equiv \sum_{I=1}^{N} m_I$ is the total mass of the system. Consequently, the multipole expansion of gravitational wave amplitudes for a regular polygonal configuration is dominated by the $l=2$, $m=0$ mode
\begin{subequations}
\begin{eqnarray}
	h_{20} = h_{20}^{+} + i h_{20}^{\times} 
	&\sim& \ddot{\mathcal{I}}_{zz} = - \frac{M}{3} \cdot \frac{d^2}{dt^2} \bigg[R^{2}(t)\bigg] , \\
	h_{21} = h_{21}^{+} + i h_{21}^{\times}
	&\sim& - (\ddot{\mathcal{I}}_{xz} - i \ddot{\mathcal{I}}_{yz}) = 0 , \\
	h_{22} = h_{22}^{+} + i h_{22}^{\times} 
	&\sim& (\ddot{\mathcal{I}}_{xx} - \ddot{\mathcal{I}}_{yy}) - 2 i \ddot{\mathcal{I}}_{xy} =0 .
\end{eqnarray}
\end{subequations}
This is in perfect agreement with the results obtained from numerical relativity in Section \ref{sec3} \footnote{Since the spin-weighted spherical harmonic $Y^{20}_{s=-2}$ is purely real, this surviving $m = 0$ mode contributes only to the plus polarization $h_{+}$, which explains why the gravitational waves emitted from the regular polygonal configurations are dominated by the plus polarization of the $l = 2$, $m = 0$ mode (i.e., $h_{20}^{(+)}$).}. The only exception is the case $N=2$, for which the coordinates of the two black holes are $x=0$ and $y=\pm R(t)$, as shown in the upper-left panel of Figure \ref{Fig BH Trajectory}. For this binary collision, the quadrupole-moment components are  $\mathcal{I}_{xx}=-\frac{MR^{2}(t)}{3}$, $\mathcal{I}_{yy}=\frac{2MR^{2}(t)}{3}$, and $\mathcal{I}_{zz}=-\frac{MR^{2}(t)}{3}$, which result in non-zero gravitational wave amplitudes for $h_{20}$ and $h_{22}$, as illustrated in the upper-left panels of Figure \ref{Fig BH Psi4} and Figure \ref{Fig BH GW Amplitude}. 

\subsection{Regular Polyhedral Configurations} \label{Appendix Polyhedron}

We now consider the regular polyhedral configurations. The following analysis suggests that gravitational radiation vanishes identically at the quadrupole approximation. For such configurations, black holes initially located at the vertices of a regular polyhedron fall radially toward the center during the collision. The position of the $I$-th black hole can therefore be written as
\begin{equation}
	\mathbf{r}_{I}(t) = R(t)\, \mathbf{n}_{I}, \qquad I = 1, 2, \dots, N, \label{nuit vector}
\end{equation}
where $R(t)$ is the common distance of all the black holes from the center, and $\mathbf{n}_{I}$ is the unit vector pointing from the center to the vertex occupied by the $I$-th black hole. Substituting equation (\ref{nuit vector}) into equation (\ref{quadrupole moment component}), we obtain the quadrupole moment
\begin{equation}
	\mathcal{I}_{ij}(t) = m R^{2}(t) \sum_{I=1}^{N} n_{Ii} \cdot n_{Ij} - \frac{1}{3} \delta_{ij} M R^{2}(t),
	\label{quadrupole moment polyhedron}
\end{equation}
where $m = m_{I} = \frac{M}{N}$ is the mass of each black hole, and $n_{Ii}$ denotes the $i$-th component of the vector $\mathbf{n}_{I}$.

Specifically, a key observation is that the vertices of any regular polyhedron (Platonic solid) --- the regular tetrahedron ($N=4$), octahedron ($N=6$), hexahedron ($N=8$), icosahedron ($N=12$), and dodecahedron ($N=20$) --- satisfy the identity
\begin{equation}
	\sum_{I=1}^{N} n_{Ii} \cdot n_{Ij} = \frac{N}{3} \cdot \delta_{ij} .
	\label{eq:polyhedron second moment}
\end{equation}
This identity can be verified in two independent ways. First, it can be verified by direct computation. For the three regular polyhedral configurations studied in Section \ref{sec3} (i.e., tetrahedron, octahedron, and hexahedron), their vertex coordinates are listed in Table \ref{Table Regular Polyhedron}. Taking the regular tetrahedron as an example, these vertex coordinates lead to the following relations for unit vectors
\begin{equation}
\begin{aligned}
		&\sum_{I=1}^{4} n_{Ix}^{2} = \sum_{I=1}^{4} n_{Iy}^{2} = \sum_{I=1}^{4} n_{Iz}^{2} = \frac{1}{3}\big(1+1+1+1\big) = \frac{4}{3} , 
		\\
		&\sum_{I=1}^{4} n_{Ix} \cdot n_{Iy} = \frac{1}{3}\big(1-1-1+1\big) = 0 , \\
		&\sum_{I=1}^{4} n_{Ix} \cdot n_{Iz} = \frac{1}{3}\big(1-1+1-1\big) = 0 , \\
		&\sum_{I=1}^{4} n_{Iy} \cdot n_{Iz} = \frac{1}{3}\big(1+1-1-1\big) = 0 ,
		\label{eq:tetrahedron sums}
\end{aligned}
\end{equation}
which is in perfect agreement with equation (\ref{eq:polyhedron second moment}). Similar computations can be carried out for the octahedron and hexahedron (cube) cases.
Second, and more generally, the vertex set of each Platonic solid is invariant under its rotation group (the tetrahedral, octahedral, or icosahedral group), and all these groups act irreducibly on $\mathbb{R}^{3}$. The tensor $S_{ij} \equiv \sum_{I} n_{Ii} \cdot n_{Ij}$ must then satisfy $S = \mathcal{R} S \mathcal{R}^{T}$ for every rotation $\mathcal{R}$ in the group. By irreducibility (Schur's lemma), the only invariant symmetric tensor is proportional to $\delta_{ij}$ \cite{Fulton1991,Dresselhaus2007}, namely $S_{ij} = \lambda \delta_{ij}$, and taking the trace fixes $\lambda = \frac{N}{3}$.

\begin{table}
	\caption{Vertex coordinates (the unit vectors $\mathbf{n}_{I}$) of the regular polyhedral configurations studied in Section \ref{sec3}.}
	\label{Table Regular Polyhedron}
	\centering
	\begin{tabular}{lccccl}
		\hline
		Configuration       & $N$ & Unit vectors $\mathbf{n}_{I}$ of the vertices \\
		\hline
		Regular Tetrahedron & 4 & $\dfrac{1}{\sqrt{3}}(1,1,1)$, $\ \dfrac{1}{\sqrt{3}}(1,-1,-1)$, $\dfrac{1}{\sqrt{3}}(-1,1,-1)$, $\ \dfrac{1}{\sqrt{3}}(-1,-1,1)$ \\
		Regular Octahedron        & 6 & $(\pm1,0,0)$, $\ (0,\pm1,0)$, $\ (0,0,\pm1)$ \\
		Regular Hexahedron (Cube) & 8 & $\dfrac{1}{\sqrt{3}}(\pm1,\pm1,\pm1)$ \\
		\hline
	\end{tabular}
\end{table}

Substituting equation (\ref{eq:polyhedron second moment}) into equation (\ref{quadrupole moment polyhedron}), we immediately obtain
\begin{equation}
	\mathcal{I}_{ij}(t) 
	= m R^{2}(t) \sum_{I=1}^{N} n_{Ii} \cdot n_{Ij} - \frac{1}{3} \delta_{ij} M R^{2}(t)
	= \frac{M R^{2}(t)}{3} \delta_{ij} - \frac{M R^{2}(t)}{3} \delta_{ij} = 0 ,
\end{equation}
that is, the quadrupole moment tensor of a black hole system in regular polyhedral configurations vanishes identically during the evolution. Consequently, from equations (\ref{h20})-(\ref{h22}), we can conclude that all the $l=2$ gravitational wave modes vanish exactly in the quadrupole approximation
\begin{equation}
	\mathcal{I}_{ij}(t) = 0 \qquad \Longrightarrow \qquad h_{20} = h_{21} = h_{22} = 0 .
\end{equation}
The gravitational radiation of a regular polyhedral configuration is therefore vanishes identically at the quadrupole level. This explains the strong suppression of all gravitational wave modes observed in the regular polyhedral simulations in Section \ref{sec3}, where the waveform amplitudes are of order $10^{-4}$ or below. The residual waveforms in Figure \ref{Fig Regular Polyhedron} therefore originate from effects beyond the quadrupole approximation (i.e., higher-order multipole contributions with $l \ge 3$), which result in very weak gravitational radiation.

\section*{Acknowledgements}

The authors gratefully acknowledge GLM-5.3 and GLM-5.3-flash for their valuable assistance in polishing the language of this paper and verifying the equations in Appendix \ref{Appendix}. This work is supported by the Natural Science Foundation of Chongqing Municipality (Grant No. CSTB2022NSCQ-MSX0932), the Scientific and Technological Research Program of Chongqing Municipal Education Commission (Grant No. KJQN202201126).

\apptocmd{\thebibliography}{\setlength{\itemsep}{0pt}}{}{}

\end{document}